\documentclass[prd,aps,floatfix]{revtex4}
\usepackage{graphicx}
\usepackage{amsmath}
\usepackage{amssymb}
\usepackage{longtable}

\newcommand{\ba}{\begin{array}}
\newcommand{\ea}{\end{array}}
\newcommand{\be}{\begin{equation}}
\newcommand{\ee}{\end{equation}}
\newcommand{\bea}{\begin{eqnarray}}
\newcommand{\eea}{\end{eqnarray}}
\newcommand{\bi}{\begin{itemize}}
\newcommand{\ei}{\end{itemize}}
\newcommand{\ben}{\begin{enumerate}}
\newcommand{\een}{\end{enumerate}}
\newcommand{\bt}{\begin{tabbing}}
\newcommand{\et}{\end{tabbing}}

\newcommand{\no}{\nonumber}

\newcommand{\csw}{c_{\rm SW}}
\newcommand{\cswnp}{c^{\rm NP}_{\rm SW}}
\newcommand{\DM}{\Delta M}
\newcommand{\PaccHMC}{P_{\rm acc}}
\newcommand{\NMD}{N_{\rm MD}}
\newcommand{\PaccGMP}{P_{\rm corr}}
\newcommand{\Npoly}{N_{\rm poly}}
\newcommand{\Ntraj}{N_{\rm traj}}
\newcommand{\gbare}{g^{2}_{0}}
\newcommand{\siml}{\rlap{\raise -4pt \hbox{$\sim$}}
                   \raise 3pt \hbox{$<$}}
\newcommand{\simg}{\rlap{\raise -4pt \hbox{$\sim$}}
                   \raise 3pt \hbox{$>$}}
\newcommand{\calO}{{\mathcal O}}

\begin{document}

\vspace*{-15mm}
\begin{flushright}
\normalsize
 KEK-CP-163 \\
 UTHEP-509 \\
 UTCCS-P-15 \\
 HUPD-0506
\end{flushright}

\title{
 Nonperturbative $O(a)$ improvement of the Wilson quark action with
 the RG-improved gauge action using the Schr\"odinger functional method
}
\author{
   S.~Aoki$^{1}$, 
   M.~Fukugita$^{2}$, 
   S.~Hashimoto$^{3,4}$, 
   K-I.~Ishikawa$^{5}$,
   N.~Ishizuka$^{1,6}$, 
   Y.~Iwasaki$^{1,6}$, 
   K.~Kanaya$^{1,6}$, 
   T.~Kaneko$^{3,4}$, 
   Y.~Kuramashi$^{1,6}$,
   M.~Okawa$^{5}$, 
   S.~Takeda$^{1}$,
   Y.~Taniguchi$^{1}$,
   N.~Tsutsui$^{3}$, 
   A.~Ukawa$^{1,6}$, 
   N.~Yamada$^{3,4}$, 
   and T.~Yoshi\'e$^{1,6}$\\
   (CP-PACS and JLQCD Collaborations)
}

\affiliation{
$^1$Institute of Physics, University of Tsukuba, Tsukuba,
    Ibaraki 305-8571, Japan \\
$^2$Institute for Cosmic Ray Research, University of Tokyo,
    Kashiwa 277-8582, Japan \\
$^3$High Energy Accelerator Research Organization(KEK), Tsukuba,
    Ibaraki 305-0801, Japan \\
$^4$The Graduate University for Advanced Studies, Tsukuba,
    Ibaraki 305-0801, Japan \\
$^5$Department of Physics, Hiroshima University, Higashi-Hiroshima,
    Hiroshima 739-8526, Japan\\
$^6$Center for Computational Sciences, University of Tsukuba, Tsukuba,
    Ibaraki 305-8577, Japan
}

\date{\today}

\begin{abstract}
 We perform a nonperturbative determination of the $O(a)$-improvement
 coefficient $c_{\rm SW}$ and the critical hopping parameter $\kappa_c$
 for $N_f$=3, 2, 0 flavor QCD with the RG-improved gauge action using the
 Schr\"odinger functional method.
 In order to interpolate $c_{\rm SW}$ and $\kappa_c$ as a function of
 the bare coupling, a wide range of $\beta$
 from the weak coupling region to the moderately strong coupling 
 points used in large-scale simulations is studied.
 Corrections at finite lattice size of $O(a/L)$ turned out to be large
 for the RG-improved gauge
 action, and hence we make the determination at a size fixed in physical
 units using a modified improvement condition.
 This enables us to avoid $O(a)$ scaling violations which would 
 remain in physical observables if $c_{\rm SW}$ determined for a fixed
 lattice size $L/a$ is
 used in numerical simulations.
\end{abstract}
\maketitle

\section{Introduction}
\label{sec:introduction}

 Fully unquenched simulations of QCD with dynamical up, down and strange
 quarks have become feasible~\cite{Ishikawa:2004nm} thanks to the recent
 development of algorithms~\cite{Kennedy:2004ae} and computational
 facilities.
 However, it is still very demanding to control discretization errors
 below a few percent level in dynamical QCD simulations.
 Thus highly improved lattice actions are desirable to accelerate
 the approach to the continuum limit.

 The on-shell improvement of the Wilson quark action through $O(a)$
 requires only a single additional term, {\it i.e.} the
 Sheikholeslami-Wohlert (SW) term~\cite{Sheikholeslami:1985ij}.
 In Ref.~\cite{Yamada:2004ja}, we determined $c_{\rm SW}$
 in three-flavor QCD for the plaquette gauge action, using
 the Schr\"odinger functional
 method~\cite{SF,NPimprovement,NPcsw.Nf0.ALPHA,NPcsw.Nf2.ALPHA}.
 Applications of the resulting $O(a)$ improved Wilson-clover quark action 
 in combination with the plaquette gauge action suffer from a serious problem, 
 however, since it was found in Ref.~\cite{Aoki:2004iq} that this
 action combination exhibits an unphysical first-order phase transition 
 at zero temperature in the strong coupling regime ($\beta\le$ 5.0). 

 We also found in Ref.~\cite{Aoki:2004iq} that such a phase
 transition weakens, and possibly disappears, when the gauge action is 
 improved.  In this work, motivated by this observation, 
 we extend the determination of $\csw$ for the case of the 
 RG-improved action~\cite{Iwasaki:1983ck} for gluons 
 for $N_f$=3, 2, 0 flavor QCD.

 We explore a wide range of $\beta$ to work out the interpolation
 formula as a function of the bare coupling.
 The critical hopping parameter $\kappa_c$ in the $O(a)$-improved theory
 is also obtained.

 In the Schr\"odinger functional method, $\csw$ is determined such that
 the axial Ward-Takahashi identity is satisfied for a given finite volume.
 Since the linear extent $L$ of a finite lattice provides an energy scale 
 $1/L$, a determination of $\csw$ generally involves 
 corrections of order $a/L$.
 We find that this correction is sizable for the RG improved 
 gauge action.  If the determination of $\csw$ is made for a fixed value of 
 $L/a$, observables calculated in subsequent simulations using such $\csw$ 
 would suffer from $O(a)$ scaling violations.
 To avoid this problem, we modify the standard improvement condition and
 determine $\csw$ at a {\it fixed physical size $L$}.
 Similar considerations have been made in the determinations of some
 other $O(a)$ improvement coefficients in
 Ref.~\cite{Luscher:1996jn,Guagnelli:2000jw}.
 
 This paper is organized as follows.
 In Sec.\ref{sec:setup}, we briefly recall the
 Schr\"odinger functional method, mainly to fix notations.
 In Sec.~\ref{sec:finite_volume}, corrections at finite lattice size of
 $O(a/L)$ that affect
 $c_{\rm SW}$ are discussed, and our modified method and one-loop
 calculations relevant for the subsequent analyses are given.
 Section~\ref{sec:simulation} is devoted to describing our numerical
 results, and Sec.~\ref{sec:sys error} to systematic uncertainties in
 them. Our conclusions are given in Sec.~\ref{sec:concl}.
 A preliminary report of this work has been made in Ref.~\cite{Aoki:2002vh}.

\section{Schr\"odinger functional method for the determination of $\csw$}
\label{sec:setup}

We briefly introduce the setup of the Schr\"odinger
functional (SF) method and the improvement condition developed in
Refs.~\cite{SF,NPimprovement,NPcsw.Nf0.ALPHA,NPcsw.Nf2.ALPHA}.

\subsection{SF setup}

Consider the SF defined on a four dimensional hypercubic lattice
with a volume $L^3\times T$ and the cylindrical geometry, {\it i.e.},
the periodic boundary condition is imposed in the spatial directions and
the Dirichlet one in the temporal direction for both gauge and quark
fields.
At the temporal boundaries $x_0=0$ and $T$,
the following conditions are imposed on the
link variables and the quark fields:
the spatial link variables on the boundaries are fixed to the diagonal,
constant $SU(3)$ matrices given by
\bea
   \left. U_k({\bf x},x_0)\right|_{x_0\!=\!0}
   = \exp \left[ aC_k \right], \ \ \ 
   \left. U_k({\bf x},x_0)\right|_{x_0\!=\!T}
   = \exp \left[ aC_k^{\prime} \right],
   \label{formul:BGF:U} \\
   C_k = \frac{i\pi}{6L_k} \left(
         \begin{array}{rrr}
            -1 & 0 & 0 \\
             0 & 0 & 0 \\
             0 & 0 & 1 
         \end{array} \right), \ \ \ 
   C_k^{\prime} = \frac{i\pi}{6L_k} \left(
         \begin{array}{rrr}
            -5 & 0 & 0 \\
             0 & 2 & 0 \\
             0 & 0 & 3 
         \end{array} \right)
   \label{formul:BGF:C},
\eea
while all quark fields on the boundaries are set to zero.

We use the RG-improved gauge action~\cite{Iwasaki:1983ck} given by,
\bea
S_g = \frac{2}{g^2}\times \Big[
      ~\sum_x~w^P_{\mu\nu}(x_0)~
       {\rm Re~Tr}\left( 1 - P_{\mu,\nu}(x) \right)~
  +   ~\sum_{x}~w^R_{\mu\nu}(x_0)~
       {\rm Re~Tr}\left( 1-R^{(1\times2)}_{\mu,\nu}(x) \right)
      \Big],
\eea
where $P_{\mu,\nu}(x)$ denotes a 1$\times$1 Wilson loop on the $\mu$-$\nu$ plane
starting and ending at $x$,
and $R^{(1\times2)}_{\mu,\nu}(x)$ a 1$\times$2 rectangular loop with the side of
length 2 in the $\nu$ direction.
These terms are added up with proper weights,
$w^P_{\mu\nu}(x_0)$ and $w^R_{\mu\nu}(x_0)$, respectively.
In ordinary simulations with the periodic boundary condition in
the temporal direction, the weights are given by
$w^P_{\mu\nu}$=3.648 and $w^R_{\mu\nu}$=$-$0.331
independently of $x_0$. In the SF, these weights are modified.
Among several possible choices, we select the choice B
defined in Ref.~\cite{Aoki:1998qd} in this work,
\bea
 w^P_{\mu\nu}(x_0) &=& \left\{\ba{ll}\displaystyle
   \frac{1}{2}\times(3.648) & \mbox{at $t=0$ or $T$ and $\mu$, $\nu\ne$4}\\[2ex]
               3.648  & \mbox{otherwise}
                  \ea\right.,\\[1ex]
w^R_{\mu\nu}(x_0) &=&  \left\{\ba{ll}
       0 & \mbox{at $t=0$ or $T$ and $\mu$, $\nu\ne$4}\\[1ex]
       \displaystyle
  \frac{3}{2}\times(-0.331) & \mbox{at $t=0$ or $T$ and $\mu$=4}\\[1ex]
              -0.331  & \mbox{otherwise}
                  \ea\right..
\eea

The $O(a)$-improved Wilson quark action~\cite{Sheikholeslami:1985ij} is
given by
\bea
     S_q
 &=& \sum_{x,y} \bar{q}_x D_{xy} q_y,
   \label{eqn:setup:Sq1}\\
     D_{xy} 
 &=& \delta_{xy} 
    -\kappa \sum_{\mu} 
     \left\{ \left( 1 - \gamma_{\mu} \right)
             U_{x,\mu} \delta_{x+\hat{\mu},y}
           + \left( 1 + \gamma_{\mu} \right)
             U_{x-\hat{\mu},\mu}^{\dagger} 
             \delta_{x-\hat{\mu},y}
     \right\}
   + \frac{i}{2} \kappa\ c_{\rm SW} \sigma_{\mu\nu} 
     F_{x,\mu\nu} \delta_{xy},
   \label{eqn:setup:Sq2}
\eea
with the field strength tensor $F_{x,\mu\nu}$
defined by 
\bea
   F_{x,\mu\nu} 
   & = & 
   \frac{1}{8} \left\{ \left( P_{ \mu, \nu}(x)
                             +P_{ \nu,-\mu}(x)
                             +P_{-\mu,-\nu}(x)
                             +P_{-\nu, \mu}(x)
                       \right)
                      -\left(\mbox{h.c.}\right)
               \right\},
   \label{eqn:setup:Fmunu}
\eea
and $\sigma_{\mu\nu} =(i/2)\left[\gamma_{\mu},\gamma_{\nu}\right]$.
The last term in Eq.~(\ref{eqn:setup:Sq2}) is the only counter term to
get rid of $O(a)$ errors present for on-shell quantities on the lattice.
At tree level, $c_{\rm SW}$=1.
For the $O(a)$-improvement of the SF, we need to add extra
terms made of the gauge and quark fields at boundaries to the
lattice action.
However, since these counter terms affect the PCAC relation used in the
following calculations only at $O(a^2)$ or higher, they are not
necessary for the determination of $c_{\rm SW}$.

\subsection{PCAC relation}

We determine $c_{\rm SW}$ by imposing the PCAC relation
\bea
   \frac{1}{2}
   \left( \partial_{\mu} + \partial_{\mu}^{*} \right)
   A_{{\rm imp},\mu}^a = 2 m_q P^a,
   \label{eq:setup:PCAC}
\eea
up to $O(a^2)$ corrections.
The pseudo-scalar density operator,
axial vector current and its $O(a)$-improved version
are given by
\bea
     P^a 
 &=& \bar{\psi} \gamma_5 \tau^a \psi,
   \label{eq:setup:P} \\
     A_{\mu}^a
 &=& \bar{\psi} \gamma_{\mu} \gamma_5 \tau^a \psi,
   \label{eq:setup:A}\\
     A_{{\rm imp},\mu}^a 
 &=& A_{\mu}^a + c_A \frac{1}{2}
     \left( \partial_{\mu} +\partial_{\mu}^{*}
     \right) P^a,
   \label{eq:setup:Aimp}
\eea     
where $\partial_{\mu}$ and $\partial_{\mu}^{*}$ are the forward and
backward lattice derivatives, and $\tau^a$ denotes the generator of
$SU(N_f)$ flavor symmetry acting on the flavor indices of the quark
fields $\bar{\psi}$ and $\psi$.

We measure two correlation functions,
\bea
   f_A(x_0) 
   & = &
   -\frac{1}{N_f^2-1} \langle A_0^a(x) \calO^a \rangle, 
   \label{eq:setup:fA}
   \\
   f_P(x_0) 
   & = &
   -\frac{1}{N_f^2-1} \langle P^a(x) \calO^a \rangle,
   \label{eq:setup:fP}
\eea
where $x=(x_0, {\bf x})$, and $\langle\cdots\rangle$ represents the
expectation value after taking trace over color and spinor indices and
summing over spatial coordinate $\bf x$.
The source operator is given by
\bea
   \calO^a
   & = &
   a^6 
   \sum_{\bf y, z} \bar{\zeta}({\bf y}) \gamma_5 
                   \tau^a \zeta({\bf z}),
   \label{eq:setup:source1}\\
   \zeta({\bf x}) 
   & = & \frac{\delta}{\delta \bar{\rho}({\bf x})}, \ \ \ 
   \bar{\zeta}({\bf x}) 
   = \frac{\delta}{\delta \rho({\bf x})},
   \label{eq:setup:zeta}
\eea
where $\rho({\bf x})$ is the quark field at $x_0\!=\!0$ and is set to
zero in the calculation of $f_A$ and $f_P$.
The bare PCAC quark mass is then calculated using $f_A$ and $f_P$
through the PCAC relation Eq.~(\ref{eq:setup:PCAC}) as
\bea
   m(x_0) & = & r(x_0) + c_A s(x_0)
   \label{eq:setup:mq}
   \\
   r(x_0) & = & \frac{1}{4}
                \left( \partial_0 + \partial_0^* \right)
                f_A(x_0) / f_P(x_0)
   \label{eq:setup:r}
   \\
   s(x_0) & = & \frac{1}{2} a \, 
                \partial_0 \partial_0^*
                f_P(x_0) / f_P(x_0).
   \label{eq:setup:s}
\eea
Using the source operator on the other boundary
\bea
   \calO^{\prime, a}
   & = &
   a^6 
   \sum_{\bf y, z} \bar{\zeta}^{\prime}({\bf y}) \gamma_5
                   \tau^a \zeta^{\prime}({\bf z}),
   \label{eq:setup:source2}
\eea
where $\zeta^{\prime}$  is the boundary field at $x_0\!=\!T$, 
we can calculate another set of quantities 
$m^{\prime}(x_0)$, $r^{\prime}(x_0)$ and
$s^{\prime}(x_0)$ from the correlation functions defined by
\bea
   f_A^{\prime}(T-x_0) 
   & = &
   +\frac{1}{N_f^2-1} \langle A_0^a(x) \calO^{\prime,a} \rangle
   \label{eq:setup:fAp},
   \\ 
   f_P^{\prime}(T-x_0) 
   & = &
   -\frac{1}{N_f^2-1} \langle P^a(x) \calO^{\prime,a} \rangle,
   \label{eq:setup:fPp}
\eea

A naive improvement condition would be $m(x_0)\!=\!m^{\prime}(x_0)$.
However, this condition requires a nonperturbative tuning of $c_A$ as
well as of $c_{\rm SW}$.
To eliminate $c_A$ from the determination, it was proposed in
Ref.~\cite{NPcsw.Nf0.ALPHA} to use an alternative definition of the quark
mass given by
\bea
     M(x_0,y_0)
 &=& m(x_0)
   - \frac{m(y_0)-m^{\prime}(y_0)}{s(y_0)-s^{\prime}(y_0)}
     s(x_0)
     \label{eq:setup:M},\\
     M'(x_0,y_0)
 &=& m'(x_0)
   - \frac{m'(y_0)-m(y_0)}{s'(y_0)-s(y_0)}
     s'(x_0)
     \label{eq:setup:M'}.
\eea
with which $c_{\rm SW}$ is obtained at the point where the mass difference
\bea 
   \Delta M(x_0,y_0) = M(x_0,y_0) - M^{\prime}(x_0,y_0)
\eea
vanishes.
In principle, we can take an arbitrary choice for $(x_0,y_0)$, since
different choices result only in $O(a^2)$ differences in physical
observables.  We follow the ALPHA Collaboration and use
$(x_0,y_0)\!=\!(3T/4,T/4)$ for $\Delta M$, and $(T/2,T/4)$ for $M$.
In the following, $M$ and $\Delta M$ without arguments denote $M(T/2,T/4)$
and $\Delta M(3T/4,T/4)$, respectively.

In previous studies, $c_{\rm SW}$ has been determined through the
conditions
\bea 
   \left\{
   \begin{array}{lll}
      M(\gbare,L/a)        & = & 0, \\
      \Delta M(\gbare,L/a) & = & \Delta M(0,L/a),
   \end{array}
   \right.
   \label{eq:setup:ImpCnd-old}
\eea
at a given $\gbare$ and $L/a$.
$\Delta M(0,L/a)$ on the right hand side, which is the tree-level
value of $\Delta M(\gbare,L/a)$ at the massless point, is necessary in
order that the resulting $\csw$ reproduces its tree-level value
($\csw$=1) in the weak coupling limit.
In the next section, we address the issue of corrections at finite
lattice size, and
propose a new condition to avoid the problem.

\section{corrections at finite lattice size and modified improvement conditions}
\label{sec:finite_volume}

\subsection{corrections at finite lattice size}
\label{subsec:finite_volume}

In the standard approach, we first calculate $M(\gbare,L/a)$ and
$\DM(\gbare,L/a)$ for a set of values of $\csw$ and $\kappa$.
The results are fitted as a function of $\csw$ and $\kappa$ to
find $\csw(\gbare,L/a)$ and
$\kappa_c(\gbare,L/a)$ satisfying Eq.~(\ref{eq:setup:ImpCnd-old}) at
a given value of $\gbare$ and $L/a$.
The asymptotic $a$ dependence of $\csw(\gbare,L/a)$ and
$\kappa_c(\gbare,L/a)$ obtained in such a way is expected to be
\begin{eqnarray}
     \csw(\gbare,L/a)
 &=& \csw(\gbare,\infty) + c_L \cdot (a/L)
   + c_\Lambda \cdot (a\Lambda_{\rm QCD})
   + O((a/L)^2,(a^2\Lambda_{\rm QCD}/L),(a\Lambda_{\rm QCD})^2),
 \label{eq:CSW_L}\\
     \kappa_c(\gbare,L/a)
 &=& \kappa_c(\gbare,\infty) + k_L \cdot (a/L)
   + k_\Lambda \cdot (a\Lambda_{\rm QCD})
   + O((a/L)^2,(a^2\Lambda_{\rm QCD}/L),(a\Lambda_{\rm QCD})^2),
 \label{eq:Kappa_L}
\end{eqnarray}
where $c_L$, $c_\Lambda$, $k_L$ and $k_\Lambda$ are unknown
coefficients.
(In practice, a logarithmic dependence on $a/L$ also  appears,
but it does not alter the following discussion, and hence not
written explicitly.)

Consider an on-shell physical quantity $Q$, and
let $Q^{\rm latt}(a)$ be the value obtained on a lattice
with lattice spacing $a$ using the SW quark action
with a choice of the improvement coefficient $\csw^{\rm sim}$.
We expect the discrepancy between $Q$ and $Q^{\rm latt}(a)$ in the
measured value to be
\begin{eqnarray}
   Q - Q^{\rm latt}(a)
 = \ q \cdot \left( \csw^{\rm sim}-\csw(\gbare,\infty)
             \right) \cdot (a\Lambda_{\rm QCD})
 + O(a^2\Lambda_{\rm QCD}^2),
    \label{eq:Q0}
\end{eqnarray}
where $q$ is an unknown constant assumed to be $O(1)$.
Hence, if one uses $\csw^{\rm sim}=\csw(\gbare,\infty)$ in the simulation,
the $O(a)$ error is absent, while if one uses
$\csw(\gbare,L/a)$ in Eq.~(\ref{eq:CSW_L}), the above expression results in
\begin{eqnarray}
   Q - Q^{\rm latt}(a)
 = q \cdot c_L \cdot  (a/L) \cdot (a\Lambda_{\rm QCD})
               + O(a^2\Lambda_{\rm QCD}^2)
               + O(a\Lambda_{\rm QCD}(a/L)^2).
    \label{eq:Q1}
\end{eqnarray}
While the scaling violation appears to start from $O(a^2)$,
{\it it is actually
linear in the lattice spacing if one determines $\csw(\gbare,L/a)$
with a fixed value of $a/L$.}
Indeed, previous studies determining $\csw$ have used certain fixed
values of $L/a$, {\it e.g.} 8, independently of $\beta$.

In Ref.~\cite{Yamada:2004ja}, we studied the magnitude of the
corrections at finite lattice size
in $\csw$ for the plaquette gauge action.  The coefficient $c_L$ defined in
Eq.~(\ref{eq:CSW_L}) was evaluated in one-loop perturbation theory in
the same SF setup, and it was found that the effect on $\csw$ does not exceed
3\% when $L/a$=8 for $\beta\ge 5.2$.
We have repeated the same perturbative analysis with the RG-improved
action, and observed a sizable effect of about 15\% at $\beta$=1.9,
around which large-scale simulations are carried out.
This enhancement of the one-loop correction for the RG improved action
is mainly due to the larger value of the bare coupling compared
to that for the plaquette gauge action for realizing the same value of
the lattice spacing.

\subsection{modified improvement condition}
\label{subsec:improvement_conditions}

We propose to resolve the problem due to the sizable corrections
explained above
by introducing a fixed physical length $L^*$, and determining $\csw$ at the
fixed physical volume ${L^*}^3\times T^*$ ($T^*=2L^*$).
If one uses $\csw$ thus determined, $L$ in (\ref{eq:Q1}) is replaced by 
$L^*$ and scaling violations are $O(a^2)$. 

The actual procedure we use runs as follows. 
Instead of Eq.~(\ref{eq:setup:ImpCnd-old}), we impose a modified 
improvement condition given by 
\begin{eqnarray}
 \left\{\ba{rrr}
   M(\gbare,L/a)&=&0,\\
 \DM(\gbare,L/a)&=&0,
  \ea\right.
 \label{eq:setup:ImpCnd-new}
\end{eqnarray}
to calculate $\csw(\gbare,L/a)$ and $\kappa_c(\gbare,L/a)$.
The results are converted to
$\csw(\gbare,L^*/a)$ and $\kappa_c(\gbare,L^*/a)$.
To do so, we must know the value of $L^*/a$ or $1/a$ at that value of
$\gbare$, which we obtain through the two-loop $\beta$ function,
\begin{eqnarray}
     a\Lambda_{L}
 &=& \exp\left( - \frac{1}{2 b_{0}g_{0}^{2}}
         \right)
         (b_{0}g_{0}^{2})^{-b_{1}/2b_{0}^{2}}
 \label{eq:scaling-2loop},\\
 b_{0} &=& \frac{1}{(4 \pi)^{2}}
           \left(\frac{11}{3} N_c -\frac{2}{3}N_f\right),\\
 b_{1} &=& \frac{1}{(4\pi)^{4}}
           \left(   \frac{34}{3} N_{c}^{2}
                  - N_f\left(   \frac{13}{3}N_c
                              - \frac{1}{N_c}
                       \right)
           \right).
\end{eqnarray}
The transformation from $\csw(\gbare,L/a)$ and $\kappa_c(\gbare,L/a)$ to
those at $L^*/a$ are made through
\begin{eqnarray}
        \csw(\gbare,L^{*}/a)
  &=&       \csw(\gbare,L/a) + \delta   \csw(\gbare,L/a;L^{*}/a),
  \label{eq:transform_csw}\\
      \kappa_c(\gbare,L^{*}/a)
  &=&     \kappa_c(\gbare,L/a) + \delta \kappa_c(\gbare,L/a;L^{*}/a),
  \label{eq:transform_kappa}
\end{eqnarray}
where
\begin{eqnarray}
        \delta\csw(\gbare,L/a;L^{*}/a)
  &=& - \csw^{\mathrm{PT}}(\gbare,L/a)
      + \csw^{\mathrm{PT}}(\gbare,L^{*}/a),
\label{eq:correction_csw}\\
        \delta \kappa_c(\gbare,L/a;L^{*}/a)
  &=& - \kappa_c^{\mathrm{PT}}(\gbare,L/a)
      + \kappa_c^{\mathrm{PT}}(\gbare,L^{*}/a),
\label{eq:correction_kc}
\end{eqnarray}
and $\csw^{\rm PT}(\gbare,L/a)$ and $\kappa_c^{\rm PT}(\gbare,L/a)$ are
calculated at the one-loop level for the same SF setup at the given value of
$L/a$.

It turned out that the tree and the one-loop coefficients for $\csw$ and
$\kappa_c$ have a significant $a/L$ dependence.
To describe this dependence precisely we fit them to a Pade or
a polynomial-like function of $a/L$ as
\begin{eqnarray}
     \csw^{(0)}(L/a)
 &=& \frac{1 + a_1\,(a/L) + a_2\,(a/L)^2 + a_3\,(a/L)^3}
          {1 + b_1\,(a/L)},
 \label{eq:TreeFinBoxCSW}\\
    \csw^{(1)}(L/a)&=& 0.113 + (c_{1}-d_{1}\ln(L/a))\,(a/L)
                             + (c_{2}-d_{2}\ln(L/a))\,(a/L)^2,
\label{eq:1LoopFinBoxCSW}\\
     \kappa_c^{(0)}(L/a)
 &=& \frac{1}{8} + k_1\, (a/L)   + k_2\,(a/L)^2
                 + k_3\, (a/L)^3 + k_4\,(a/L)^4,
\label{eq:TreeFinBoxKC}\\
\kappa_c^{(1)}(L/a)&=& 0.002760894 + (l_{1}-m_{1}\ln(L/a))\,(a/L)
                                   + (l_{2}-m_{2}\ln(L/a))\,(a/L)^2
\label{eq:1LoopFinBoxKC}.
\end{eqnarray}
The coefficients are given in Table~\ref{tab:FinBoxCSWKC}.
We note that the one-loop coefficients have an $N_f$ dependence due to the
tadpole diagram, although it vanishes in the large volume limit.

In our actual determination, we define $L^*$ by $L^*/a=6$ at $\beta=1.9$,
$L^*/a=6$ at $\beta=2.0$ and $L^*/a=6$ at $\beta=2.6$ for $N_f$=3, 2, 0
flavor QCD, respectively.
In Table~\ref{tab:Nf3latsize6}--\ref{tab:Nf0latsize6} numerical values of
$\beta=6/\gbare$, $L/a$ and $L^*/a$ in our simulations for $N_f$=3, 2, and 0
cases are summarized.
In these tables, we also show the numerical values of
$\delta\csw(\gbare,L/a;L^{*}/a)$ and
$\delta\kappa_c(\gbare,L/a;L^{*}/a)$.
For large values of $\beta$, the perturbative corrections are small and hence
reliable.
On the other hand, if $L/a$ are close to $L^*/a$, the corrections
needed for the conversion from $L$ to $L^*$ should
again be small.
Since we fix $L^*$ at strong coupling,
the corrections, Eqs.~(\ref{eq:correction_csw}) and
(\ref{eq:correction_kc}), are small at both ends of our range of $\beta$
as one can see in the Tables.

\section{Numerical simulations}
\label{sec:simulation}

\subsection{parameters and algorithm}

The numerical simulations are performed with $N_f$=3, 2 and 0 degenerate
dynamical quarks on a $(L/a)^3 \times 2(L/a)$ ($L/a$= 8 or 6) lattice
for a wide range of $\beta$.
The simulation parameters are summarized in
Tabs.~\ref{tab:Nf3latsize6}--\ref{tab:Nf0latsize6} for $N_f$=3, 2, 0,
respectively.

We employ the symmetric even-odd preconditioning introduced in
Refs.~\cite{Even-Odd,PHMC.JLQCD} for the quark matrix $D$.
Calculation of $D^{-1}$ is made with the BiCGStab algorithm with the tolerance
parameter $||R_i||/||B|| \! < \! 10^{-14}$, where $R_i=DX_i-B$ is the
residual vector and $X_i$ is an estimate for the solution $X$ in the
$i$-th BiCGStab iteration.

We adopt the standard HMC algorithm~\cite{HMC} for the $N_f$=2 and 0
flavor cases.  For the three-flavor case, the polynomial HMC
(PHMC) algorithm~\cite{PHMC,PHMC.JLQCD} is applied to
describe the third flavor,
employing the Chebyshev polynomial $P[D]$ to approximate $D^{-1}$.
In order to make the PHMC algorithm exact, the correction factor
$P_{\rm corr}={\rm det}[W[D]]$ with $W[D]\!=\!P[D]D$ is taken into
account by the noisy Metropolis method~\cite{NoisyMetroP}.
The square root of $W[D]$, which is required in the
Metropolis test, is evaluated with an accuracy of $10^{-14}$ using a Taylor
expansion of $W[D]$~\cite{PHMC.JLQCD}.
The order of the polynomial $N_{\rm poly}$ is chosen so that an
acceptance rate of about 70\% or higher is achieved for the Metropolis test.

In the calculations of $aM$ and $a\DM$, $f_X$ and $f_X'$ ($X$=$A$ or
$P$) are first evaluated at every trajectory, and they are combined to
produce $aM$ and $a\DM$.
The bin size dependence of the jackknife error of $aM$
is investigated in the range
$N_{\rm bin}\!=\!1$\,--\,$N_{\rm traj}/20$.
We adopt $N_{\rm bin}$ giving the maximum error in this range
in the error analyses in the following.

\subsection{results}
\label{subsec:results}

The trial values of $\csw$ and $\kappa$ at which simulations are made
are summarized in Tables.~\ref{tab:Nf3MandDM}--\ref{tab:Nf0MandDM} for
$N_f$=3, 2, and 0, respectively, together with the results for $aM$
and $a\DM$ and the number of trajectories accumulated.
In order to obtain $\csw(\gbare,L/a)$ and $\kappa_c(\gbare,L/a)$ satisfying
Eq.~(\ref{eq:setup:ImpCnd-new}) at each $\beta$, we make
fits of those data using the functional forms,
\bea
      a M
  &=& a_M + \frac{b_{M}^{(1)}}{\kappa} + \frac{b_{M}^{(2)}}{\kappa^2}
    + c_M^{(1)} \, c_{\rm SW} + c_M^{(2)} \, {c_{\rm SW}}^2
    + \frac{d_M}{\kappa}\, c_{\rm SW},
   \label{eq:Nf3:combfit:M}
   \\
      a \Delta M
  &=& a_{\Delta M} + \frac{b_{\Delta M}^{(1)}}{\kappa}
    + \frac{b_{\Delta M}^{(2)}}{\kappa^2}
    + c_{\Delta M}^{(1)} \, \csw
    + c_{\Delta M}^{(2)} \, \csw^2
    + \frac{d_{\Delta M}}{\kappa}\, \csw.
   \label{eq:Nf3:combfit:dM}
\eea
The results for $\csw(\gbare,L/a)$ and $\kappa_c(\gbare,L/a)$ obtained 
with the fits, and the adopted functional form are tabulated in
Tabs.~\ref{tab:Nf3NPTKCSW}--\ref{tab:Nf0NPTKCSW}.
The details of the fit procedure are as follows.
In Figs.~\ref{fig:MandDM_3f}--\ref{fig:MandDM_0f}
we plot data on the $(aM,a\DM)$ plane for $N_f$=3, 2, 0,
respectively.
For those data for which the origin $(0,0)$ is contained in or close 
to the data region, we make a fit leaving only the constant and linear 
terms in Eqs.~(\ref{eq:Nf3:combfit:M}) and (\ref{eq:Nf3:combfit:dM}). 
This applies to all cases 
except for the three-flavor simulations at $\beta\le$ 2.2, 
and the dotted lines in the figures show the fit results.

In the three-flavor simulations at $\beta\le$ 2.2,
the region of negative $aM$ is not covered, and the origin is missed
by the data. 
This happens because the PHMC algorithm tends to fail at vanishing or
negative PCAC quark masses at low $\beta$ due to large quantum
fluctuations.
Thus, at $\beta\le$ 2.2, we are forced to extrapolate the data.
In the extrapolation, three functional forms are examined:
(i) linear, (ii) quadratic without the cross terms, and (iii)
quadratic with the cross terms.
At $\beta$=2.20 and 2.10, a linear function well fits the data, and
we take this in the following analysis.
The data at $\beta$=2.00 and 1.90 require the quadratic term, but it
turns out that including the cross terms does not reduce $\chi^2$/dof
significantly from that without the cross terms, and leads to
$\csw(\gbare,L/a)$ and $\kappa_c(\gbare,L/a)$ consistent within one
standard deviation.
Thus, we adopt the quadratic function without the cross terms at these
$\beta$, and $d_M$ and $d_{\Delta M}$ are always set to zero throughout
this analysis.

Next, $\csw(\gbare,L/a)$ and $\kappa_c(\gbare,L/a)$ are transformed into
those for the desired lattice volume, $(L^*/a)^3\times 2(L^*/a)$, along
the line presented in Sec.~\ref{subsec:improvement_conditions}.
Using Eqs.~(\ref{eq:transform_csw}), (\ref{eq:transform_kappa}) and the
$\delta\csw$ and $\delta\kappa_c$ given in
Tables~\ref{tab:Nf3latsize6}--\ref{tab:Nf0latsize6}, we obtain
$\csw(\gbare,L^*/a)$ and $\kappa_c(\gbare,L^*/a)$ shown in
Tables~\ref{tab:Nf3NPTKCSW_b19_l6}--~\ref{tab:Nf0NPTKCSW_b26_l6}.
Notice that in Table~\ref{tab:Nf3NPTKCSW_b19_l6} there are three results
for $\beta$=2.0.
The first and second one are obtained by transforming the data with
$8^3\times16$ and $6^3\times12$ to those for $L^*/a\sim$6.805,
respectively, and the third one is obtained by simply interpolating the
two raw values at $L/a=8$ and $6$ in Table~\ref{tab:Nf3NPTKCSW} to
$L^*/a\sim$6.805, for which the corrections at finite lattice size are
essentially
corrected nonperturbatively.
The two raw values, 1.670(56) at $L/a=8$ and 1.632(45) at $L/a=6$, are
very close to each other and consistent within the error, and hence the
linear interpolation to $L^*/a\sim 6.805$ is more reliable than
the perturbative procedure.
Similar observations are made at the second smallest $\beta$ in
each $N_f$ flavor simulation, namely at
$\beta$=2.10 for $N_f=2$ and at $\beta$=2.70 for $N_f=0$.
Thus, at these $\beta$ the result interpolated to $L^*/a$ is adopted as
our final result, and used in the following analysis.
At the same time, it is worth noting that in all three cases the
one-loop corrections have the right sign,
which indicates that the one-loop correction dominates over higher loop
corrections.
Furthermore, the discrepancy between the results corrected
perturbatively and nonperturbatively is found to be 5\%, 3\% and less
than 1\% for the $N_f$=3, 2 and 0 cases, respectively, while the size of
one-loop correction itself at these $\beta$ is 6--7\%, 5\% and 2--3\%.
From this observation, we expect that the size of the one-loop
correction gives a conservative estimate for the unknown higher loop
corrections for all $\beta$.

\subsection{interpolation formula}
\label{subsec:interpolation}

Our final results for $\csw(\gbare,L^*)$ as a function of $\gbare$ 
are shown in Fig.~\ref{fig:Nf320Csw} for $N_f=3, 2, 0$ flavor QCD.
When we interpolate $\csw$, not all available data are used in the fit.
As mentioned in Sec.~\ref{subsec:improvement_conditions}, 
the corrections at finite lattice size estimated perturbatively is
small only around the high and low ends of $\beta$ due to our
choice of $L^*$, while in the middle range corrections may be significant.
Therefore, we use data only if the
correction is less than 5\%.
In the three flavor case, the data at $\beta$=12.0, 8.85, 2.2, 2.1, 2.0,
1.9 are employed.
As a consequence, we obtain the followings interpolation formula,
\begin{eqnarray}
      \csw(\gbare,L^*)
  &=& 1 + 0.113\ \gbare + 0.0209(72)\ (\gbare)^{2}
        + 0.0047(27)\ (\gbare)^{3},\ \ \ \
 (\chi^2/{\rm dof} = 0.58).
\label{eq:Nf3cswformula}
\end{eqnarray}
For $\kappa_c$ shown in Fig.~\ref{fig:Nf320KappaC},
the corrections are smaller than 5\% for all value of $\beta$.
Including all data in the fit we obtain
\begin{eqnarray}
      \kappa_c(\gbare,L^*)
  &=& 1/8 + 0.003681192\ \gbare
          + 0.000211(43)\ (\gbare)^{2}\no\\
  & &\ \ \ \ \ \ 
          + 0.000067(66)\ (\gbare)^{3}
          - 0.000038(21)\ (\gbare)^{4}.
     \ \ \ \
  (\chi^2/{\rm dof}=1.1)
\label{eq:Nf3kappacformula}
\end{eqnarray}
When performing the above fits, the tree and one-loop coefficients
are fixed to the perturbative values at infinite volume.
This is justified since, as seen in Table~\ref{tab:Nf3latsize6},
$L^*/a$ grows very rapidly with $\beta$, and hence $a/L^*$ corrections in
Eqs.~(\ref{eq:TreeFinBoxCSW})--(\ref{eq:1LoopFinBoxKC}) are all
negligibly small near the continuum limit.
We also note that the tree and one-loop coefficients in the infinite volume
limit do not depend on $N_f$, and hence
the same values are used in the analysis for the $N_f$= 2 and 0 cases
given below.

The interpolation formula for $\csw$ in two-flavor QCD is calculated in
the same fashion as in the three-flavor case.
In this case, the sizes of the correction at finite lattice size are
acceptable ($\le$ 5\%) at $\beta$=12.0, 8.85, 5,0, 2.2, 2.1, 2.0.
We first try a polynomial form as before, and obtain 
\begin{eqnarray}
     \csw(\gbare,L^*)
 &=& 1 + 0.113\,\gbare + 0.0158(63)\,(\gbare)^{2}
       + 0.0088(24)\,(\gbare)^{3},\ \ \
 (\chi^2/{\rm dof}=4.68),
\label{eq:Nf2cswformula_poly}
\end{eqnarray}
which is denoted by a dashed line in Fig.~\ref{fig:Nf320Csw}.
A sharp rise of the data points near $\gbare$=3.0 is not described well
by this polynomial form, while in the three-flavor case the polynomial
worked well over the whole range of $\beta$ we studied.
An alternative is a Pade function, with which we obtain
\begin{eqnarray}
     \csw(\gbare,L^*)
 &=& \frac{1 -0.212(9)\,\gbare -0.0108(38)\,(\gbare)^2-0.0083(19)\,(\gbare)^3}
          {1 -0.325(9)\,\gbare},\ \ \
 (\chi^2/{\rm dof}=2.11),
\label{eq:Nf2cswformula}
\end{eqnarray}
This fit, denoted by solid line in the middle panel of
Fig.~\ref{fig:Nf320Csw}, interpolates our data very well.
Since this formula has a pole at $\gbare$=3.08(8), its use is restricted
to $\gbare\siml$3.0.
For $\kappa_c$, we use all available data to obtain
\begin{eqnarray}
     \kappa_c(\gbare,L^*)
 &=& 1/8 + 0.003681192\,\gbare
         + 0.000227(58)\,(\gbare)^2\no\\
 & &\ \ \ \ \ \
         + 0.000093(84)\,(\gbare)^3 
         - 0.000049(24)\,(\gbare)^4,\ \ \
 (\chi^2/{\rm dof}=0.98),
\label{eq:Nf2kappacformula_poly}
\end{eqnarray}
for a polynomial, and
\begin{eqnarray}
     \kappa_c(\gbare,L^*)
 &=& \frac{1/8 - 0.0356(23)\,\gbare -0.00089(8)\,(\gbare)^2
          - 0.00009(6)\,(\gbare)^3}
          {1 - 0.314(18)\,\gbare},\ \ \
 (\chi^2/{\rm dof}=0.35),
\label{eq:Nf2kappacformula}
\end{eqnarray}
for a Pade function.
These results appear in the middle panel of Fig.~\ref{fig:Nf320KappaC}
as dashed and solid line, respectively.
It is interesting that the pole positions for $\csw$ and $\kappa_c$ are
consistent with each other.
This seems to indicate that above $\gbare\sim$3.0 the Wilson quark
action cannot be improved in this fashion consistently for the $N_f$=2
case.
All in all the Pade fits provide a more satisfactory interpolation of
the $N_f$=2 data, and we take them as the main result for the $N_f$=2
case.
We have also applied a Pade function for $\csw$ in the $N_f$=3 case.
However, in this case the resulting fit lies on top of that for a
polynomial over the range of $\beta$ we used, and the position of pole
can be determined only poorly.
Hence there seems no reason to favor the Pade fit over the polynomial
for interpolating the data.
The difference between the $N_f$=2 and 3 cases probably arise from the
fact, empirically known, that the $N_f$=2 lattice is coarser than the
$N_f$=3 lattice at the same value of $\gbare$.
Indeed, a sharp rise of improvement coefficients was previously seen for
the plaquette gauge action toward coarse
lattices~\cite{NPcsw.Nf0.ALPHA,NPcsw.Nf2.ALPHA}.

In quenched QCD, the size of the correction is smaller
than 5\% for all availble data, and we use all data to obtain
\begin{eqnarray}
\csw(\gbare,L^*)&=&1+0.113\,\gbare+0.0371(54)\,(\gbare)^2
                    -0.0036(26)\,(\gbare)^3,\ \ \ \
(\chi^2/{\rm dof}=4.09),
\label{eq:Nf0cswformula}\\
     \kappa_{c}(\gbare,L^*)
 &=& 1/8 + 0.003681192\,\gbare
         + 0.000293(37)\,(\gbare)^2\no\\
 & &\ \ \ \ \ \
         - 0.000053(65)\,(\gbare)^3
         + 0.000008(24)\,(\gbare)^4,\ \ \
    (\chi^2/{\rm dof}=0.46).
\label{eq:Nf0kappacformula}
\end{eqnarray}


In Ref.~\cite{Aoki:2003sj}, the authors performed a one-loop
determination of $\csw^{(1)}$ with conventional perturbation theory,
and reported a very precise value $\csw^{(1)}$=0.11300591(1) in the
infinite volume limit.  Changes in our results due to the use of this value
in above analyses are expected to be negligibly small.

\section{Systematic errors}
\label{sec:sys error}

There are two sources of systematic errors in our analysis, both related to
the conversion to a fixed physical length scale $L^*$, one being the
use of the two-loop $\beta$ function to estimate $L^*$ as a function of
$\gbare$, and the second being the use of one-loop perturbation theory
for correcting the value of $\csw$ from $L$ to $L^*$.

In order to examine the magnitude of uncertainties from the first error,
we go through the analysis using the three-loop $\beta$ function.
Since the three-loop term of the lattice $\beta$ function is not available
for the RG-improved gauge action, we take the value for the plaquette gauge
action. Thus the following argument is only semi-quantitatively valid.
In this case, Eq.~(\ref{eq:scaling-2loop}) is replaced with
\begin{eqnarray}
     a\Lambda_{L}
 &=& \exp\left( - \frac{1}{2 b_{0}g_{0}^{2}}
         \right)
         (b_{0}\gbare)^{-b_{1}/2b_{0}^{2}}
  \times \left(1+q \gbare\right)
 \label{eq:scaling-3loop},
\end{eqnarray}
where $q$=0.18960350(1), 0.4529(1), and 0.6138(2) for $N_f$=0, 2, and
3~\cite{Bode:2001uz}, respectively.
With this function, we estimate $L^*/a$, $\delta\csw$, and
$\delta\kappa_c$ with $N_f$=3, which are tabulated in
Table~\ref{tab:sys error 1}.
Comparing with Table~\ref{tab:Nf3latsize6}, it is found that $L^*/a$
changes significantly while the changes in $\delta\csw$ and
$\delta\kappa_c$ are at most a few percent and hence small.
Thus we conclude that the uncertainty from scaling violation in the
lattice spacing is negligible.

In order to discuss the uncertainty of one-loop corrections,
we write $\csw(\gbare,L^*)$ determined
through our procedure as
\begin{eqnarray}
     \csw(\gbare,L^*)
 &=& \csw(\gbare,\infty)
   + c^{(0)}\,(a/L^*) +
     g_0^2\,c^{(1)}(a/L^*) + g_0^4\,c^{(2)}(a/L^*)
\no\\&&
   + g_0^4\left(\,c^{(2)}(a/L)-c^{(2)}(a/L^*)\right)
   + O(g_0^6).
   \label{eq:csw error}
\end{eqnarray}
In other words, Eq.~(\ref{eq:csw error}) represents the difference
between $\csw(\gbare,L^*)$ and $\csw(\gbare,\infty)$ in terms of
perturbative series with coefficients $c^{(i)}(a/L)$, where
$c^{(i)}(a/L)$ vanishes as $L\rightarrow\infty$.
Since we have corrected the mismatch between $\csw(\gbare,L/a)$ and
$\csw(\gbare,L^*/a)$ only at the tree- and 
one-loop level, the unwanted $a/L$ dependence remains at two-loop and
higher.
Replacing $\csw^{\rm sim}$ in Eq.~(\ref{eq:Q0}) with
Eq.~(\ref{eq:csw error}), we obtain
\begin{eqnarray}
 \hspace*{-3ex}
    Q - Q^{\rm latt}(a)
&=& \bigg(
      c^{(0)}(a/L^*) + \gbare\,c^{(1)}(a/L^*) + g_0^4\,c^{(2)}(a/L^*)
    \bigg)\cdot(a\Lambda_{\rm QCD})\no\\&& \hspace*{-1ex}
  + g_0^4\,\big(c^{(2)}(a/L) - c^{(2)}(a/L^*) \big)\cdot
    (a\Lambda_{\rm QCD})
  + O(g_0^6\,a\Lambda_{\rm QCD}\,a/L)
  + O(a^2\Lambda_{\rm QCD}^2),
  \label{eq:error in phys}
\end{eqnarray}
where we omit an unknown $O(1)$ overall coefficient $q$, because it is
not relevant in the following discussion.
If you expand $c^{(i)}(a/L^{(*)})$ around $a/L^{(*)}=0$, the first
term in Eq.~(\ref{eq:error in phys}) behaves
$\sim a^2\Lambda_{\rm QCD}/L^*\sim O(a^2)$ because $L^*$ is fixed.
The second term behaves like $\sim g_0^4(a/L-a/L^*)(a\Lambda_{\rm QCD})$,
which gives $O(a)$ scaling violation because $a/L$ is fixed.
As a results, the leading scaling violation could be $O(a)$ rather than
$O(a^2)$.
However it should be emphasized that when we obtain the interpolation
formula we only used the weak coupling and strong coupling regions
because in these regions the perturbative errors are expected to be
under control for the following reasons.
In the weak coupling region, $L/a$ and $L^*/a$ are different by several
orders of magnitude, but the coupling is very small, and hence the
size of $O(g^4(a/L-a/L^*)(a \Lambda_{\rm QCD}))$ is expected to be as
small as the size of the one-loop corrections.
On the other hand, in the strong coupling region, $L/a$ and $L^*/a$ are
close to each other, and again the remaining $O(a)$ scaling violation,
$O(g^4(a/L-a/L^*)(a \Lambda_{\rm QCD}))$, should be small.
We also saw in Sec.~\ref{subsec:results}
that the size of perturbative errors is roughly the same as that of the
one-loop correction itself.

Most importantly, at our strongest and the second strongest couplings around
which large-scale simulations are performed, there are no perturbative
errors in $\csw$ due to our choice of $L^*$ and interpolation to $L^*$
at the second strongest couplings.
Thus we believe $O(a)$ scaling violations are well below $O(a^2)$,
though we need to check this in future work.

\section{Conclusion}
\label{sec:concl}

In this work, we have performed a nonperturbative determination of the
$O(a)$-improvement coefficient $\csw$ of the Wilson quark action with
the RG-improved gauge action for $N_f$=3, 2, and 0 flavor QCD.
The corrections at finite lattice size turn out to be sizable, and are taken into
account by modifying the improvement condition and carrying out the
determination at a fixed physical length scale of $L^*$.
While we have to resort to perturbation theory to incorporate the
corrections, we have attempted to
choose $L^*$ at a moderately strong coupling, close to the range of
lattice sizes of order $a^{-1}\sim 2$~GeV where physics simulations are
practically made, so that their magnitude are reasonably under control.

Using the data for $\csw$ thus obtained over a wide range of $\beta$, we have
determined the interpolation formulas, given in
Eqs.~(\ref{eq:Nf3cswformula}), (\ref{eq:Nf2cswformula}) and
(\ref{eq:Nf0cswformula}), which represent the main results of this work.
These results do depend on $L^*$ chosen, but the removal of
$O(a)$ scaling violations in physical observables hold
independent of the value of $L^*$.

As a byproduct, we have also obtained the interpolation formula for
$\kappa_c$, Eqs.~(\ref{eq:Nf3kappacformula}),
(\ref{eq:Nf2kappacformula}) and (\ref{eq:Nf0kappacformula}), which may
be useful to locate simulation points.

The three-flavor results reported here are already being used in a 
large-scale simulation aiming to carry out a systematic evaluation of 
hadronic observables for the realistic quark spectrum incorporating 
the dynamical up, down and strange quarks.  
The preliminary results have been reported in Refs.~\cite{Kaneko:2003re}.

\begin{acknowledgments}
This work is supported by the Supercomputer Project No.132 (FY2005)
of High Energy Accelerator Research Organization (KEK),
and also in part by the Grant-in-Aid of the Ministry of Education
(Nos. 13135204, 14740173, 15204015, 15540251, 16028201, 16540228,
 17340066, 17540259).
\end{acknowledgments}


\clearpage
\begin{table}
\centering
\caption{Finite-size coefficients in
 Eqs.~(\ref{eq:TreeFinBoxCSW})--(\ref{eq:1LoopFinBoxKC}).}
\label{tab:FinBoxCSWKC}
  \begin{tabular}{cr||crrr}
\multicolumn{2}{c||}{\smash{\lower 6pt \hbox{$\csw^{(0)}$}}} &
\multicolumn{4}{c}{$\csw^{(1)}$}\\
&&& $N_f=0$ & $N_f=2$ & $N_f=3$ \\\hline
  $a_1$ &$-$3.4415
& $c_1$ &$-$4.5736 & $-$6.2641 & $-$7.1094 \\
  $a_2$ &$-$5.0248
& $c_2$ &$-$3.3402 & $-$8.0488 & $-$10.403 \\
  $a_3$ &  11.1475
& $d_1$ &$-$1.1681 & $-$1.5466 & $-$1.7359 \\
  $b_1$ &$-$3.9702
& $d_2$ &$-$8.9448 & $-$14.306 & $-$16.987 \\\hline
\multicolumn{2}{c||}{\smash{\lower 7pt \hbox{$\kappa_c^{(0)}$}}} &
\multicolumn{4}{c}{$\kappa_c^{(1)}$}\\
&&& $N_f=0$ & $N_f=2$ & $N_f=3$ \\\hline
 $k_1$&\  0.260982$\times10^{-6}$ & $l_1$ & 0.101302$\times10^{-2}$
      &$-$0.224650$\times10^{-2}$ &      $-$0.387626$\times10^{-2}$ \\
 $k_2$&$-$0.845333$\times10^{-5}$ & $l_2$ & 0.162496$\times10^{-1}$
      &\  0.862878$\times10^{-2}$ &         0.481835$\times10^{-2}$ \\
 $k_3$&$-$0.103610$\times10^{-1}$ & $m_1$ & 0.547826$\times10^{-3}$
      &$-$0.507665$\times10^{-3}$ &      $-$0.155835$\times10^{-3}$ \\
 $k_4$&\  0.751742$\times10^{-2}$ & $m_2$ & 0.882220$\times10^{-2}$
      &$-$0.136413$\times10^{-2}$ &      $-$0.645729$\times10^{-2}$ \\
  \end{tabular}
\end{table}
\begin{table}
\centering
\caption{Inverse coupling $\beta$ and lattice size $L/a$ chosen for
 the three-flavor QCD simulation.
 $L^*/a$ is estimated by the two-loop $\beta$ function assuming
 $L^{*}/a=6$ at $\beta$=1.9.
 Finite-size corrections $\delta\csw$ and $\delta\kappa_c$ calculated
 with Eqs.(\ref{eq:correction_csw}) and(\ref{eq:correction_kc}) are also
 shown.}
\label{tab:Nf3latsize6}
  \begin{ruledtabular}
  \begin{tabular}{rclrr}
$\beta$ & $L/a$ &  $L^{*}/a$
        & $\delta     \csw(\gbare,L/a;L^{*}/a)$ 
        & $\delta \kappa_c(\gbare,L/a;L^{*}/a)$\\\hline
 12.00 & 8 &   7.51$\times 10^{6}$
           &   5.51$\times10^{-3}$ &   6.35$\times10^{-5}$\\
  8.85 & 8 &   8.46$\times 10^{4}$
           &   1.42$\times10^{-2}$ &   7.95$\times10^{-5}$\\
  5.00 & 8 &   3.81$\times 10^{2}$
           &   5.14$\times10^{-2}$ &   1.23$\times10^{-4}$\\
  3.00 & 8 &   2.50$\times 10^{1}$
           &   1.14$\times10^{-1}$ &   6.80$\times10^{-5}$\\
  2.60 & 8 &   1.48$\times 10^{1}$
           &   1.08$\times10^{-1}$ &   1.34$\times10^{-5}$\\
  2.40 & 8 &   1.14$\times 10^{1}$
           &   8.70$\times10^{-2}$ &$-$8.82$\times10^{-6}$\\
  2.20 & 8 &   8.78
           &   3.42$\times10^{-2}$ &$-$9.84$\times10^{-6}$\\
  2.10 & 8 &   7.73
           &$-$1.59$\times10^{-2}$ &   5.70$\times10^{-6}$\\
  2.00 & 8 &   6.81
           &$-$9.36$\times10^{-2}$ &   3.85$\times10^{-5}$\\
  2.00 & 6 &   6.81
           &   1.10$\times10^{-1}$ &$-$5.08$\times10^{-5}$\\
  1.90 & 6 &  6 &  0 &  0\\
  \end{tabular}
  \end{ruledtabular}
\end{table}
\begin{table}
\centering                
\caption{Same as Table~\ref{tab:Nf3latsize6}, but for two-flavor QCD.}
\label{tab:Nf2latsize6}
  \begin{ruledtabular}
  \begin{tabular}{rclrr}
$\beta$ & $L/a$ &  $L^{*}/a$
        & $\delta   \csw(L/a;L^{*}/a)$ 
        & $\delta \kappa(L/a;L^{*}/a)$\\\hline
 12.00 & 8 & 2.35$\times 10^{ 6}$ &   2.43$\times10^{-3}$ &   5.93$\times10^{-5}$\\
  8.85 & 8 & 3.66$\times 10^{ 4}$ &   1.01$\times10^{-2}$ &   7.38$\times10^{-5}$\\
  5.00 & 8 & 2.45$\times 10^{ 2}$ &   4.63$\times10^{-2}$ &   1.08$\times10^{-4}$\\
  3.00 & 8 & 1.98$\times 10^{ 1}$ &   9.51$\times10^{-2}$ &   2.69$\times10^{-5}$\\
  2.60 & 8 & 1.22$\times 10^{ 1}$ &   7.84$\times10^{-2}$ &$-$1.81$\times10^{-5}$\\
  2.20 & 8 & 7.58                 &$-$2.11$\times10^{-2}$ &   1.41$\times10^{-5}$\\
  2.10 & 8 & 6.74                 &$-$8.24$\times10^{-2}$ &   6.14$\times10^{-5}$\\
  2.10 & 6 & 6.74                 &   8.37$\times10^{-2}$ &$-$7.04$\times10^{-5}$\\
  2.00 & 6 & 6                    &   0                   &   0\\
  \end{tabular}
  \end{ruledtabular}
\end{table}
\begin{table}
\centering
\caption{Same as Table~\ref{tab:Nf3latsize6}, but for quenched QCD.}
\label{tab:Nf0latsize6}
  \begin{ruledtabular}
  \begin{tabular}{rclrr}
$\beta$ & $L/a$ &  $L^{*}/a$
        & $\delta   \csw(L/a;L^{*}/a)$ 
        & $\delta \kappa(L/a;L^{*}/a)$\\\hline
 24.00 & 8 & 3.09$\times 10^{11}$&$-$1.11$\times10^{-2}$ &   3.47$\times10^{-5}$\\
 12.00 & 8 & 2.41$\times 10^{ 5}$&$-$3.70$\times10^{-3}$ &   5.08$\times10^{-5}$\\
  8.85 & 8 & 6.33$\times 10^{ 3}$&   2.24$\times10^{-3}$ &   6.18$\times10^{-5}$\\
  5.00 & 8 & 8.04$\times 10^{ 1}$&   3.80$\times10^{-2}$ &   6.29$\times10^{-5}$\\
  3.00 & 8 & 9.12                &   2.07$\times10^{-2}$ &$-$2.68$\times10^{-5}$\\
  2.70 & 8 & 6.66                &$-$4.81$\times10^{-2}$ &   8.79$\times10^{-5}$\\
  2.70 & 6 & 6.66                &   3.98$\times10^{-2}$ &$-$8.40$\times10^{-5}$\\
  2.60 & 6 & 6                   &   0                   &   0 \\
  \end{tabular}
  \end{ruledtabular}
 \vspace{-3ex}
\end{table}
\clearpage
\begin{longtable}{ccccccc}
\caption{Results for $aM$ and $a\DM$ for three-flavor QCD.
 The acceptance rates for the MD and the noisy Metropolis test are shown
 together with the number of MD steps per trajectory and the order
 of the polynomial $N_{\rm poly}$ used in the noisy Metropolis test. 
 The final column gives the number of trajetories accumulated.}
\endfirsthead
\label{tab:Nf3MandDM}
\endhead
\hline\hline
\endlastfoot
\endfoot
\hline\hline
$\csw$&$\kappa$&$aM$&$a\DM$&$\PaccHMC[\NMD]$&$\PaccGMP[\Npoly]$&$\Ntraj$
 \\\hline\\[1ex]
  \multicolumn{7}{c}{$\beta= 12.00$, $L/a= 8$}\\[1ex] 
 1.00 & 0.12659 & 0.01235(13) & 0.00101(13) & 0.73(2)[100] & 0.983(5)[100]& 1600 \\
      & 0.12676 & 0.006906(91)& 0.00072(13) & 0.75(1)[100] & 0.979(4)[100]& 1600 \\
      & 0.12693 & 0.00149(13) & 0.00087(14) & 0.73(2)[100] & 0.973(5)[100]& 1600 \\
      & 0.12709 &$-$0.00368(13) & 0.00092(16) & 0.75(1)[100] & 0.970(5)[100]& 1600 \\
 1.05 & 0.12659 & 0.008565(98)&$-$0.00009(18) & 0.72(2)[100] & 0.984(4)[100]& 1600 \\
      & 0.12676 & 0.00283(11) &$-$0.00003(13) & 0.74(1)[100] & 0.968(6)[100]& 1600 \\
      & 0.12693 &$-$0.00221(11) & 0.00004(17) & 0.72(3)[100] & 0.969(6)[100]& 1600 \\
      & 0.12709 &$-$0.007708(89)& 0.00023(10) & 0.74(2)[100] & 0.955(5)[100]& 1600 \\
 1.10 & 0.12659 & 0.00460(12) &$-$0.00076(13) & 0.72(2)[100] & 0.981(4)[100]& 1600 \\
      & 0.12676 &$-$0.00097(15) &$-$0.00069(22) & 0.71(3)[100] & 0.972(4)[100]& 1600 \\
      & 0.12693 &$-$0.00625(21) &$-$0.00073(16) & 0.74(2)[100] & 0.953(9)[100]& 1600 \\
      & 0.12709 &$-$0.01161(17) &$-$0.00070(13) & 0.73(2)[100] & 0.94(2)[100] & 1600 \\
  \hline\hline\\[1ex]
  \multicolumn{7}{c}{$\beta= 8.85$, $L/a= 8$}\\[1ex]
 1.0141 & 0.12698 & 0.02316(13) & 0.00094(15) & 0.68(2)[80] & 0.989(3)[100] & 2000 \\
        & 0.12730 & 0.01311(11) & 0.00104(18) & 0.66(2)[80] & 0.988(4)[100] & 2000 \\
        & 0.12762 & 0.00309(12) & 0.00102(15) & 0.71(2)[80] & 0.969(5)[100] & 2000 \\
        & 0.12826 &$-$0.01734(10) & 0.00092(16) & 0.70(1)[80] & 0.943(6)[110] & 2000 \\
 1.0350 & 0.12698 & 0.02121(11) & 0.00060(12) & 0.70(2)[80] & 0.990(2)[100] & 2000 \\
        & 0.12730 & 0.01119(16) & 0.00057(17) & 0.71(1)[80] & 0.985(3)[100] & 2000 \\
        & 0.12762 & 0.00080(13) & 0.00060(16) & 0.70(2)[80] & 0.972(4)[100] & 2000 \\
        & 0.12826 &$-$0.01948(13) & 0.00055(15) & 0.71(2)[80] & 0.937(6)[110] & 2000 \\
 1.0559 & 0.12698 & 0.01903(14) & 0.00013(14) & 0.69(2)[80] & 0.987(4)[100] & 2000 \\
        & 0.12730 & 0.00896(11) & 0.00033(10) & 0.68(1)[80] & 0.975(4)[100] & 2000 \\
        & 0.12762 &$-$0.00130(11) & 0.00024(20) & 0.69(2)[80] & 0.964(5)[100] & 2000 \\
        & 0.12826 &$-$0.02166(13) & 0.00039(10) & 0.70(2)[80] & 0.928(7)[110] & 2000 \\
 1.0800 & 0.12719 & 0.00983(15) &$-$0.00013(15) & 0.69(2)[80] & 0.990(2)[120] & 2000 \\
        & 0.12753 &$-$0.00078(11) & 0.00007(12) & 0.68(2)[80] & 0.989(3)[130] & 2000 \\
 1.1000 & 0.12713 & 0.00982(35) &$-$0.00041(19) & 0.68(2)[80] & 0.990(3)[110] & 2000 \\
        & 0.12747 &$-$0.00121(11) &$-$0.00039(14) & 0.69(1)[80] & 0.986(3)[120] & 2000 \\
  \hline\hline\\[1ex]
  \multicolumn{7}{c}{$\beta= 5.00$, $L/a= 8$}\\[1ex]
 1.08 & 0.12958 & 0.01031(33) & 0.00073(22) & 0.72(1)[64] & 0.982(3)[100] & 2200 \\
      & 0.12974 & 0.00553(17) & 0.00082(29) & 0.77(2)[64] & 0.968(4)[100] & 2200 \\
      & 0.12989 & 0.00049(35) & 0.00060(25) & 0.74(2)[64] & 0.970(4)[100] & 2200 \\
      & 0.13004 &$-$0.00377(20) & 0.00070(19) & 0.74(2)[64] & 0.962(5)[100] & 2200 \\
 1.13 & 0.12932 & 0.01027(26) & 0.00039(22) & 0.73(1)[64] & 0.976(6)[100] & 2200 \\
      & 0.12948 & 0.00541(24) & 0.00043(27) & 0.73(1)[64] & 0.975(4)[100] & 2200 \\
      & 0.12963 & 0.00030(24) & 0.00024(21) & 0.77(1)[64] & 0.964(7)[100] & 2200 \\
      & 0.12978 &$-$0.00428(29) & 0.00042(37) & 0.74(1)[64] & 0.950(6)[100] & 2200 \\
 1.18 & 0.12907 & 0.01002(22) &$-$0.00071(20) & 0.76(2)[64] & 0.982(3)[100] & 2200 \\
      & 0.12922 & 0.00489(25) &$-$0.00105(22) & 0.73(2)[64] & 0.974(4)[100] & 2200 \\
      & 0.12937 & 0.00043(43) &$-$0.00074(21) & 0.75(1)[64] & 0.970(4)[100] & 2200 \\
      & 0.12952 &$-$0.00441(28) &$-$0.00089(21) & 0.76(1)[64] & 0.958(6)[100] & 2200 \\
  \hline\hline\\[1ex]
  \multicolumn{7}{c}{$\beta= 3.00$, $L/a= 8$}\\[1ex]
 1.20 & 0.13281 & 0.02798(33) & 0.00099(26) & 0.770(9)[50]& 0.988(2)[100] & 4300 \\
      & 0.13311 & 0.01769(47) & 0.00036(57) & 0.77(1)[50] & 0.978(3)[100] & 4000 \\
      & 0.13341 & 0.00813(46) & 0.00044(33) & 0.78(1)[50] & 0.960(4)[100] & 4000 \\
      & 0.13370 &$-$0.00036(41) & 0.00104(40) & 0.77(1)[50] & 0.937(5)[100] & 3800 \\
 1.25 & 0.13235 & 0.02774(42) &$-$0.00015(61) & 0.78(1)[50] & 0.986(4)[100] & 4200 \\
      & 0.13265 & 0.01820(52) & 0.00003(40) & 0.76(1)[50] & 0.980(3)[100] & 3800 \\
      & 0.13294 & 0.00940(38) & 0.00004(36) & 0.77(1)[50] & 0.960(3)[100] & 4200 \\
      & 0.13324 &$-$0.00027(40) &$-$0.00029(34) & 0.773(9)[50]& 0.948(5)[100] & 3900 \\
 1.30 & 0.13190 & 0.02742(65) &$-$0.00052(28) & 0.76(1)[50] & 0.990(2)[100] & 4200 \\
      & 0.13219 & 0.01713(64) &$-$0.00040(36) & 0.77(2)[50] & 0.980(2)[100] & 4200 \\
      & 0.13248 & 0.00915(54) &$-$0.00066(67) & 0.78(2)[50] & 0.962(3)[100] & 3900 \\
      & 0.13278 &$-$0.00008(41) & 0.00031(64) & 0.77(1)[50] & 0.945(6)[100] & 4000 \\
 1.35 & 0.13145 & 0.02697(35) &$-$0.00098(46) & 0.77(1)[50] & 0.988(3)[100] & 4300 \\
      & 0.13174 & 0.01643(70) &$-$0.00075(44) & 0.770(8)[50]& 0.979(4)[100] & 4100 \\
      & 0.13203 & 0.00800(43) &$-$0.00092(58) & 0.76(1)[50] & 0.972(3)[100] & 4100 \\
      & 0.13232 &$-$0.00079(38) &$-$0.00109(38) & 0.77(1)[50] & 0.947(4)[100] & 4000 \\
  \hline\hline\\[1ex]
  \multicolumn{7}{c}{$\beta= 2.60$, $L/a= 8$}\\[1ex]
 1.20 & 0.13531 & 0.02110(64) & 0.00168(38) & 0.878(9)[64] & 0.979(3)[110] & 4500 \\
      & 0.13550 & 0.01528(44) & 0.00142(85) & 0.879(6)[64] & 0.972(3)[110] & 4500 \\
      & 0.13574 & 0.00810(69) & 0.00170(43) & 0.87(1)[64]  & 0.966(6)[120] & 4500 \\
      & 0.13594 & 0.00140(72) & 0.00158(62) & 0.882(7)[64] & 0.965(8)[130] & 4500 \\
 1.27 & 0.13454 & 0.02061(53) & 0.00073(91) & 0.870(6)[64] & 0.983(2)[110] & 4500 \\
      & 0.13473 & 0.01512(73) & 0.00218(47) & 0.881(6)[64] & 0.978(2)[110] & 4500 \\
      & 0.13496 & 0.00721(52) & 0.00102(48) & 0.883(10)[64]& 0.971(5)[120] & 4500 \\
      & 0.13516 & 0.00327(73) & 0.00039(53) & 0.883(6)[64] & 0.972(3)[130] & 4500 \\
 1.34 & 0.13378 & 0.02177(75) &$-$0.00006(57) & 0.883(6)[64] & 0.984(2)[110] & 4500 \\
      & 0.13420 & 0.00830(57) &$-$0.00004(45) & 0.872(6)[64] & 0.972(3)[120] & 4500 \\
      & 0.13440 & 0.0018(11)  & 0.00031(44) & 0.876(9)[64] & 0.975(2)[130] & 4500 \\
      & 0.13473 &$-$0.00968(93) &$-$0.00007(34) & 0.880(6)[64] & 0.903(5)[110] & 4500 \\
 1.41 & 0.13303 & 0.02040(50) &$-$0.00055(43) & 0.874(7)[64] & 0.983(2)[110] & 4500 \\
      & 0.13322 & 0.01375(64) &$-$0.00042(67) & 0.874(7)[64] & 0.979(3)[110] & 4500 \\
      & 0.13344 & 0.00792(81) &$-$0.00064(37) & 0.873(9)[64] & 0.975(3)[120] & 4500 \\
      & 0.13364 & 0.00116(66) &$-$0.00022(50) & 0.882(6)[64] & 0.964(3)[130] & 4500 \\
 1.48 & 0.13277 &  0.0036(10) &$-$0.00109(49) & 0.883(6)[64] & 0.978(3)[130] & 5000 \\
      & 0.13301 &$-$0.00316(79) &$-$0.00085(34) & 0.872(6)[64] & 0.979(2)[150] & 5000 \\
 1.55 & 0.13202 & 0.00476(58) &$-$0.00218(36) & 0.873(7)[64] & 0.980(2)[130] & 5000 \\
      & 0.13226 &$-$0.00270(73) &$-$0.00195(48) & 0.879(10)[64]& 0.973(3)[140] & 5000 \\
  \hline\hline\\[1ex]
  \multicolumn{7}{c}{$\beta= 2.40$, $L/a= 8$}\\[1ex]
 1.3 & 0.135917 & 0.0211(39) & 0.00162(74) & 0.819(7)[50] & 0.974(2)[110] & 10000 \\
     & 0.136152 & 0.01146(50)& 0.00011(100)& 0.819(5)[50] & 0.971(2)[120] & 10000 \\
     & 0.136387 & 0.00276(49)& 0.00100(29) & 0.82(1)[50]  & 0.947(2)[120] & 10000 \\
 1.4 & 0.134882 & 0.01207(50)& 0.00022(60) & 0.815(5)[50] & 0.974(2)[120] & 10000 \\
     & 0.135113 & 0.00466(45)& 0.00034(50) & 0.828(5)[50] & 0.954(2)[120] & 10000 \\
 1.5 & 0.133410 & 0.0206(10) &$-$0.00040(30) & 0.823(4)[50] & 0.983(2)[110] & 10000 \\
     & 0.133636 & 0.01300(56)& 0.00027(68) & 0.828(5)[50] & 0.979(2)[120] & 10000 \\
     & 0.133862 & 0.00584(49)&$-$0.00095(39) & 0.827(5)[50] & 0.966(2)[120] & 10000 \\
 1.6 & 0.132400 & 0.0151(12) &$-$0.00210(81) & 0.82(1)[50]  & 0.980(2)[120] & 11900 \\
     & 0.132680 & 0.00567(70)&$-$0.00194(55) & 0.825(5)[50] & 0.966(2)[120] & 11900 \\
 1.7 & 0.131230 & 0.01385(71)&$-$0.00236(54) & 0.888(5)[64] & 0.983(1)[120] & 10700 \\
     & 0.131510 & 0.0047(13) &$-$0.00321(28) & 0.886(7)[64] & 0.978(1)[130] & 10700 \\
  \hline\hline\\[1ex]
  \multicolumn{7}{c}{$\beta= 2.20$, $L/a= 8$}\\[1ex]
 1.3 & 0.138247 & 0.01685(90) & 0.00161(63) & 0.840(5)[50] & 0.958(2)[120] & 16500 \\
     & 0.138487 & 0.0100(15)  & 0.00144(36) & 0.836(3)[50] & 0.946(3)[130] & 16500 \\
     & 0.138729 & 0.00128(62) & 0.00149(45) & 0.843(4)[50] & 0.919(3)[140] & 16500 \\
 1.5 & 0.135400 & 0.01877(64) &$-$0.00028(38) & 0.844(4)[50] & 0.973(1)[120] & 16000 \\
     & 0.135654 & 0.01083(46) &$-$0.00004(36) & 0.844(4)[50] & 0.965(2)[130] & 16500 \\
     & 0.135885 & 0.00285(55) & 0.00079(51) & 0.841(4)[50] & 0.951(2)[140] & 16500 \\
 1.7 & 0.132712 & 0.01913(70) &$-$0.00148(29) & 0.846(5)[50] & 0.983(1)[120] & 16500 \\
     & 0.132934 & 0.01226(45) &$-$0.00198(45) & 0.844(8)[50] & 0.977(2)[130] & 16500 \\
     & 0.133156 & 0.00494(49) &$-$0.00205(50) & 0.834(5)[50] & 0.968(2)[140] & 16500 \\
 1.9 & 0.130170 & 0.02045(54) &$-$0.00385(32) & 0.843(6)[50] & 0.984(1)[120] & 16100 \\
     & 0.130370 & 0.01248(76) &$-$0.00342(30) & 0.840(4)[50] & 0.984(2)[130] & 16100 \\
     & 0.130570 & 0.00561(55) &$-$0.00322(32) & 0.841(4)[50] & 0.977(1)[140] & 16100 \\
  \hline\hline\\[1ex]
  \multicolumn{7}{c}{$\beta= 2.10$, $L/a= 8$}\\[1ex]
 1.5 & 0.1355 & 0.0579(12) & 0.00066(57) & 0.850(4)[50] & 0.979(3)[80]   & 14500 \\
     & 0.1358 & 0.0500(15) & 0.00011(44) & 0.847(4)[50] & 0.9877(10)[100]& 14500 \\
     & 0.1360 & 0.04380(77)& 0.00033(44) & 0.855(4)[50] & 0.984(1)[100]  & 14400 \\
     & 0.1362 & 0.0386(10) & 0.00047(56) & 0.851(6)[50] & 0.978(2)[100]  & 14500 \\
 1.6 & 0.1340 & 0.06167(75)&$-$0.00027(49) & 0.854(5)[50] & 0.986(2)[80]   & 14500 \\
     & 0.1344 & 0.04811(59)& 0.00039(83) & 0.848(5)[50] & 0.9888(10)[100]& 14500 \\
     & 0.1346 & 0.04179(72)&$-$0.00035(76) & 0.850(4)[50] & 0.985(1)[100]  & 14500 \\
 1.7 & 0.1326 & 0.05964(58)&$-$0.00132(47) & 0.853(5)[50] & 0.987(1)[80]   & 14500 \\
     & 0.1329 & 0.04952(50)&$-$0.00128(33) & 0.850(4)[50] & 0.987(2)[90]   & 14500 \\
     & 0.1331 & 0.04393(48)&$-$0.00145(32) & 0.854(3)[50] & 0.988(1)[100]  & 14500 \\
     & 0.1333 & 0.03752(54)&$-$0.00018(31) & 0.849(3)[50] & 0.982(2)[100]  & 14500 \\
     & 0.1335 & 0.02976(87)&$-$0.00096(31) & 0.856(3)[50] & 0.974(1)[100]  & 14500 \\
 1.8 & 0.1315 & 0.05054(44)&$-$0.00192(33) & 0.850(4)[50] & 0.9889(9)[90]  & 14500 \\
     & 0.1318 & 0.04101(59)&$-$0.00182(35) & 0.844(3)[50] & 0.9883(10)[100]& 14500 \\
     & 0.1321 & 0.03149(47)&$-$0.00176(41) & 0.847(4)[50] & 0.983(2)[110]  & 14500 \\
     & 0.1324 & 0.01956(59)&$-$0.00148(33) & 0.850(4)[50] & 0.980(1)[120]  & 14500 \\
  \hline\hline\\[1ex]
  \multicolumn{7}{c}{$\beta= 2.00$, $L/a= 8$}\\[1ex]
 1.5 & 0.1383550 & 0.0276(15) & 0.00020(44) & 0.863(3)[50] & 0.961(2)[130] & 24500 \\
     & 0.1386672 & 0.0176(12) & 0.00082(56) & 0.857(3)[50] & 0.928(2)[140] & 21500 \\
     & 0.1388000 & 0.0138(13) & 0.00034(58) & 0.853(3)[50] & 0.914(7)[160] & 21200 \\
 1.6 & 0.1364310 & 0.0360(10) & 0.00036(60) & 0.858(4)[50] & 0.9912(9)[140]& 20000 \\
     & 0.1367500 & 0.02618(98)&$-$0.0011(10)  & 0.857(4)[50] & 0.982(1)[150] & 20000 \\
     & 0.1370700 & 0.0156(12) & 0.00049(52) & 0.857(7)[50] & 0.961(2)[160] & 20000 \\
 1.7 & 0.1348740 & 0.03723(61)&$-$0.00134(41) & 0.856(3)[50] & 0.9932(9)[140]& 20000 \\
     & 0.1354309 & 0.01588(70)&$-$0.00073(30) & 0.861(3)[50] & 0.950(2)[130] & 24500 \\
     & 0.1354980 & 0.01435(93)&$-$0.00037(46) & 0.859(3)[50] & 0.971(1)[160] & 20000 \\
     & 0.1356000 & 0.01043(82)&$-$0.00072(58) & 0.861(4)[50] & 0.951(2)[150] & 20300 \\
 1.8 & 0.1333520 & 0.03533(61)&$-$0.00095(40) & 0.859(3)[50] & 0.9936(6)[140]& 20000 \\
     & 0.1336570 & 0.02441(57)&$-$0.00129(33) & 0.863(3)[50] & 0.9899(8)[150]& 20000 \\
     & 0.1339620 & 0.01374(65)&$-$0.00050(51) & 0.853(5)[50] & 0.978(1)[160] & 20000 \\
 1.9 & 0.1320578 & 0.02843(57)&$-$0.00237(38) & 0.860(3)[50] & 0.977(1)[110] & 24500 \\
     & 0.1323422 & 0.01918(68)&$-$0.00166(42) & 0.860(2)[50] & 0.976(1)[130] & 24500 \\
     & 0.1326278 & 0.01016(60)&$-$0.00183(39) & 0.858(3)[50] & 0.945(2)[130] & 24500 \\
 2.0 & 0.1308300 & 0.0208(13) &$-$0.00357(75) & 0.853(6)[50] & 0.989(2)[150] & 5400 \\
     & 0.1311100 & 0.01208(82)&$-$0.00310(40) & 0.860(7)[50] & 0.983(2)[160] & 7300 \\
 2.1 & 0.1293800 & 0.02589(84)&$-$0.00391(65) & 0.859(8)[50] & 0.996(2)[150] & 5400 \\
     & 0.1296600 & 0.01483(96)&$-$0.00406(79) & 0.855(5)[50] & 0.988(2)[160] & 7300 \\
  \hline\hline\\[1ex]
  \multicolumn{7}{c}{$\beta= 2.00$, $L/a= 6$}\\[1ex]
 1.30 & 0.1400 & 0.1002(43) & 0.0055(15) & 0.925(3)[50] & 0.9998(2)[120]  & 11200 \\
      & 0.1400 & 0.1017(33) & 0.00346(89) & 0.926(3)[50] & 1.0000(0)[120] & 11200 \\
      & 0.1405 & 0.0871(30) & 0.00390(95) & 0.923(2)[50] & 0.9993(2)[120] & 14500 \\
      & 0.1405 & 0.0890(29) & 0.0030(11)  & 0.926(2)[50] & 0.9991(5)[120] & 14500 \\
      & 0.1410 & 0.0669(28) & 0.0033(11)  & 0.922(3)[50] & 0.9974(6)[120] & 15000 \\
      & 0.1410 & 0.0721(30) & 0.0045(15)  & 0.922(2)[50] & 0.9974(8)[120] & 15000 \\
      & 0.1415 & 0.0505(26) & 0.0037(11)  & 0.922(4)[50] & 0.990(1)[120]  & 15100 \\
      & 0.1415 & 0.0508(24) & 0.0048(12)  & 0.923(2)[50] & 0.990(1)[120]  & 15100 \\
 1.45 & 0.1380 & 0.0752(22) & 0.00153(95) & 0.925(3)[50] & 0.9995(2)[120] & 14900 \\
      & 0.1380 & 0.0784(17) & 0.00126(66) & 0.925(3)[50] & 0.9994(3)[120] & 14900 \\
      & 0.1385 & 0.0564(17) & 0.00224(73) & 0.921(3)[50] & 0.9985(4)[120] & 14900 \\
      & 0.1385 & 0.0568(24) & 0.00094(76) & 0.922(4)[50] & 0.9979(5)[120] & 14900 \\
      & 0.1390 & 0.0401(20) & $-$0.0001(12) & 0.921(2)[50] & 0.994(1)[120]  & 15000 \\
      & 0.1390 & 0.0417(21) & 0.00173(83) & 0.921(3)[50] & 0.993(1)[120]  & 15000 \\
      & 0.1395 & 0.0231(17) & 0.00128(80) & 0.923(5)[50] & 0.982(2)[120]  & 15100 \\
      & 0.1395 & 0.0245(23) & 0.00238(83) & 0.926(3)[50] & 0.981(1)[120]  & 15100 \\
 1.60 & 0.1355 & 0.0673(20) & 0.0021(16)  & 0.925(4)[50] & 0.9994(2)[120] & 11200 \\
      & 0.1355 & 0.0698(16) &$-$0.00090(78) & 0.924(3)[50] & 0.9998(2)[120] & 11200 \\
      & 0.1360 & 0.0530(17) &$-$0.00053(56) & 0.920(3)[50] & 0.9983(6)[120] & 15000 \\
      & 0.1360 & 0.0535(14) & $-$0.0004(14) & 0.925(3)[50] & 0.9990(3)[120] & 15000 \\
      & 0.1365 & 0.0340(19) & 0.0008(10)  & 0.926(2)[50] & 0.9957(8)[120] & 15000 \\
      & 0.1365 & 0.0347(15) & 0.00078(62) & 0.925(3)[50] & 0.9971(5)[120] & 15000 \\
      & 0.1370 & 0.0165(14) &$-$0.00013(69) & 0.922(3)[50] & 0.988(1)[120]  & 15000 \\
      & 0.1370 & 0.0171(16) & $-$0.0004(12) & 0.923(2)[50] & 0.9888(10)[120]& 15000 \\
 1.75 & 0.1330 & 0.0682(37) &$-$0.00223(95) & 0.920(7)[50] & 1.0000(0)[120] & 3400 \\
      & 0.1330 & 0.0683(20) &$-$0.0016(10)  & 0.927(7)[50] & 0.9997(3)[120] & 3400 \\
      & 0.1335 & 0.05035(96)&$-$0.00293(66) & 0.925(3)[50] & 0.9996(3)[120] & 14500 \\
      & 0.1335 & 0.05242(96)&$-$0.00187(64) & 0.923(3)[50] & 0.9995(2)[120] & 14500 \\
      & 0.1340 & 0.0339(10) &$-$0.00088(59) & 0.927(3)[50] & 0.9985(5)[120] & 15000 \\
      & 0.1340 & 0.0350(11) & $-$0.0025(13) & 0.924(3)[50] & 0.9983(5)[120] & 15000 \\
      & 0.1345 & 0.0164(15) & $-$0.0026(10) & 0.920(3)[50] & 0.9948(6)[120] & 15000 \\
      & 0.1345 & 0.0197(15) &$-$0.00095(70) & 0.920(3)[50] & 0.994(1)[120]  & 15000 \\
 \hline\hline\\[1ex]
 \multicolumn{7}{c}{$\beta= 1.90$, $L/a= 6$}\\[1ex]
 1.4 & 0.1410 & 0.1315(62) & 0.00112(92) & 0.930(2)[50] & 0.9997(1)[120] & 24000 \\
     & 0.1410 & 0.1338(41) & 0.0008(13)  & 0.923(2)[50] & 0.9997(1)[120] & 23900 \\
     & 0.1415 & 0.1103(64) & 0.0006(12)  & 0.928(2)[50] & 0.9985(4)[120] & 24000 \\
     & 0.1415 & 0.1136(36) & 0.00144(94) & 0.927(2)[50] & 0.9989(2)[120] & 24000 \\
     & 0.1420 & 0.0857(25) & 0.0028(10)  & 0.930(3)[50] & 0.9930(8)[120] & 24100 \\
     & 0.1420 & 0.0876(31) & 0.00166(90) & 0.923(2)[50] & 0.9945(5)[120] & 24100 \\
 1.8 & 0.1340 & 0.0878(25) &$-$0.00021(84) & 0.929(3)[50] & 1.0000(0)[120] & 15100 \\
     & 0.1340 & 0.0900(21) &$-$0.00174(58) & 0.928(3)[50] & 1.0000(0)[120] & 15100 \\
     & 0.1345 & 0.0710(21) &$-$0.00159(69) & 0.926(3)[50] & 0.9994(2)[120] & 24000 \\
     & 0.1345 & 0.0733(14) &$-$0.00193(55) & 0.929(2)[50] & 0.9995(3)[120] & 24000 \\
     & 0.1350 & 0.0520(24) &$-$0.00136(59) & 0.929(2)[50] & 0.9973(3)[120] & 24100 \\
     & 0.1350 & 0.0535(21) &$-$0.00045(76) & 0.928(2)[50] & 0.9976(6)[120] & 24100 \\
     & 0.1355 & 0.0307(16) &$-$0.00102(68) & 0.926(2)[50] & 0.9888(7)[120] & 24100 \\
     & 0.1355 & 0.0309(14) &$-$0.00108(55) & 0.926(2)[50] & 0.9895(8)[120] & 24100 \\
 2.2 & 0.1280 & 0.06146(84)&$-$0.00601(53) & 0.928(2)[50] & 0.99991(9)[120]& 22700 \\
     & 0.1280 & 0.06303(92)&$-$0.00699(67) & 0.926(2)[50] & 0.9998(1)[120] & 22700 \\
     & 0.1285 & 0.04376(79)&$-$0.00734(52) & 0.928(2)[50] & 0.9993(2)[120] & 23600 \\
     & 0.1285 & 0.04434(83)&$-$0.00594(51) & 0.928(2)[50] & 0.9996(1)[120] & 23600 \\
     & 0.1290 & 0.02540(87)&$-$0.00681(84) & 0.929(2)[50] & 0.9976(6)[120] & 24100 \\
     & 0.1290 & 0.02607(99)&$-$0.00585(67) & 0.924(2)[50] & 0.9976(4)[120] & 24100 \\
     & 0.1295 & 0.00747(99)& $-$0.0044(16) & 0.925(2)[50] & 0.9916(8)[120] & 24100 \\
     & 0.1295 & 0.0084(15) &$-$0.00649(46) & 0.928(2)[50] & 0.9922(6)[120] & 24100 \\
 2.5 & 0.1240 & 0.0535(12) &$-$0.01090(60) & 0.926(2)[50] & 0.99990(7)[120]& 22200 \\
     & 0.1240 & 0.0539(12) &$-$0.01076(50) & 0.927(2)[50] & 0.9997(2)[120] & 22200 \\
     & 0.1245 & 0.0372(13) &$-$0.01083(49) & 0.929(2)[50] & 0.9996(1)[120] & 24000 \\
     & 0.1250 & 0.01913(94)&$-$0.01152(87) & 0.926(2)[50] & 0.9985(3)[120] & 24000 \\
     & 0.1250 & 0.0198(11) &$-$0.01144(38) & 0.925(2)[50] & 0.9986(3)[120] & 24100 \\
     & 0.1255 & 0.00201(98)&$-$0.01160(49) & 0.924(2)[50] & 0.9937(7)[120] & 24100 \\
     & 0.1255 &$-$0.00014(87)&$-$0.01100(44) & 0.923(2)[50] & 0.9932(8)[120] & 24100 \\
\end{longtable}
\begin{longtable*}{cccccc}
\caption[]{Same as Table~\ref{tab:Nf3MandDM} for two-flavor QCD.}\\
\endfirsthead
\caption[]{(\textit{Continued})}
\label{tab:Nf2MandDM}
\endhead
\endlastfoot
\endfoot
\hline\hline
 $\csw$ & $\kappa$ & $aM$  & $a\DM$ &$\PaccHMC[\NMD]$& $\Ntraj$
 \\\hline\\[1ex]
 \multicolumn{6}{c}{$\beta= 12.00$, $L/a= 8$}\\[1ex]
 1.00 & 0.12659 & 0.01266(15) & 0.00082(16) & 0.75(3)[100] & 1500 \\
      & 0.12676 & 0.007137(77)& 0.00083(14) & 0.74(2)[100] & 1500 \\
      & 0.12693 & 0.00180(16) & 0.00115(16) & 0.73(2)[100] & 1500 \\
      & 0.12709 &$-$0.003231(98)& 0.00070(13) & 0.75(2)[100] & 1500 \\
 1.05 & 0.12659 & 0.00882(14) & 0.00003(20) & 0.71(3)[100] & 1500 \\
      & 0.12676 & 0.00317(11) & 0.00019(14) & 0.73(2)[100] & 1500 \\
      & 0.12693 &$-$0.00199(11) & 0.00016(14) & 0.70(2)[100] & 1500 \\
      & 0.12709 &$-$0.00735(16) &$-$0.00008(12) & 0.72(2)[100] & 1500 \\
 1.10 & 0.12659 & 0.00498(12) &$-$0.00061(11) & 0.74(2)[100] & 1500 \\
      & 0.12676 &$-$0.00042(12) &$-$0.00065(15) & 0.73(1)[100] & 1500 \\
      & 0.12693 &$-$0.00581(11) &$-$0.00069(12) & 0.71(2)[100] & 1500 \\
      & 0.12709 &$-$0.01119(15) &$-$0.00070(12) & 0.74(2)[100] & 1500 \\
 \hline\hline\\[1ex]
 \multicolumn{6}{c}{$\beta= 8.85$, $L/a= 8$}\\[1ex]
 1.040 & 0.1270 & 0.020524(91)& 0.00036(11) & 0.72(2)[80] & 2100 \\
       & 0.1274 & 0.00794(13) & 0.00056(11) & 0.72(2)[80] & 2100 \\
       & 0.1278 &$-$0.00470(14) & 0.00052(15) & 0.70(1)[80] & 2100 \\
       & 0.1282 &$-$0.017306(94)& 0.00043(14) & 0.70(1)[80] & 2000 \\
 1.055 & 0.1270 & 0.019165(83)& 0.00044(14) & 0.70(2)[80] & 2000 \\
       & 0.1274 & 0.00665(14) & 0.00042(16) & 0.70(2)[80] & 2000 \\
       & 0.1278 &$-$0.00628(12) & 0.00032(18) & 0.67(2)[80] & 2000 \\
       & 0.1282 &$-$0.01900(12) & 0.00025(16) & 0.70(2)[80] & 2000 \\
 1.070 & 0.1270 & 0.01758(10) & 0.00002(15) & 0.70(1)[80] & 2000 \\
       & 0.1274 & 0.00505(12) & 0.00001(18) & 0.70(1)[80] & 2000 \\
       & 0.1278 &$-$0.00754(14) & 0.00027(14) & 0.69(2)[80] & 2000 \\
       & 0.1282 &$-$0.02057(17) & 0.00012(11) & 0.70(2)[80] & 2000 \\
 \hline\hline\\[1ex]
 \multicolumn{6}{c}{$\beta= 5.00$, $L/a= 8$}\\[1ex]
 1.09 & 0.12954 & 0.01204(16) & 0.00054(26) & 0.75(1)[64] & 2300 \\
      & 0.12970 & 0.00692(23) & 0.00068(21) & 0.76(2)[64] & 2300 \\
      & 0.12986 & 0.00198(25) & 0.00041(21) & 0.74(1)[64] & 2300 \\
      & 0.13002 &$-$0.00308(18) & 0.00086(43) & 0.74(1)[64] & 2300 \\
 1.13 & 0.12933 & 0.01167(21) &$-$0.00005(31) & 0.74(2)[64] & 2300 \\
      & 0.12949 & 0.00691(21) &$-$0.00014(24) & 0.75(1)[64] & 2300 \\
      & 0.12965 & 0.00174(15) & 0.00041(19) & 0.75(1)[64] & 2300 \\
      & 0.12981 &$-$0.00307(24) &$-$0.00013(20) & 0.75(1)[64] & 2300 \\
 1.17 & 0.12912 & 0.01164(26) &$-$0.00063(18) & 0.76(2)[64] & 2300 \\
      & 0.12928 & 0.00667(27) &$-$0.00031(28) & 0.74(2)[64] & 2300 \\
      & 0.12943 & 0.00175(32) &$-$0.00076(20) & 0.75(1)[64] & 2300 \\
      & 0.12959 &$-$0.00338(19) &$-$0.00020(33) & 0.75(1)[64] & 2300 \\
 \hline\hline\\[1ex]
 \multicolumn{6}{c}{$\beta= 3.00$, $L/a= 8$}\\[1ex]
 1.20 & 0.1332100 & 0.02195(25) & 0.00069(38) & 0.780(5)[50] & 10500 \\
      & 0.1333700 & 0.01674(36) & 0.00086(22) & 0.767(7)[50] & 10500 \\
      & 0.1335400 & 0.01165(34) & 0.00044(28) & 0.77(1)[50]  & 10500 \\
      & 0.1337000 & 0.00685(22) & 0.00068(28) & 0.762(5)[50] & 10500 \\
 1.28 & 0.1324700 & 0.02175(35) &$-$0.00013(29) & 0.774(6)[50] & 10500 \\
      & 0.1326300 & 0.01668(33) &$-$0.00007(21) & 0.773(6)[50] & 10500 \\
      & 0.1327900 & 0.01168(30) &$-$0.00008(40) & 0.781(8)[50] & 10500 \\
      & 0.1329500 & 0.00649(35) & 0.00021(22) & 0.766(5)[50] & 10500 \\
 1.36 & 0.1317300 & 0.02146(39) &$-$0.00079(21) & 0.769(7)[50] & 10500 \\
      & 0.1318900 & 0.01644(30) &$-$0.00068(24) & 0.771(7)[50] & 10500 \\
      & 0.1320400 & 0.01156(37) &$-$0.00099(28) & 0.777(7)[50] & 10500 \\
      & 0.1322000 & 0.00675(30) &$-$0.00092(22) & 0.779(6)[50] & 10500 \\
 \hline\hline\\[1ex]
 \multicolumn{6}{c}{$\beta= 2.60$, $L/a= 8$}\\[1ex]
 1.20 & 0.135574 & 0.02473(74) & 0.00104(62) & 0.809(7)[50] & 4500 \\
      & 0.135738 & 0.0191(12)  & 0.00161(59) & 0.800(8)[50] & 4500 \\
      & 0.135903 & 0.01522(75) & 0.00209(53) & 0.808(7)[50] & 4500 \\
      & 0.136068 & 0.01002(58) & 0.00143(93) & 0.806(8)[50] & 4500 \\
 1.25 & 0.135020 & 0.02424(81) & 0.0015(10)  & 0.80(1)[50]  & 4500 \\
      & 0.135180 & 0.01912(58) & 0.00149(49) & 0.811(8)[50] & 4500 \\
      & 0.135340 & 0.01506(67) & 0.00156(56) & 0.810(8)[50] & 4500 \\
      & 0.135510 & 0.00982(59) & 0.00150(82) & 0.812(7)[50] & 4500 \\
 1.30 & 0.134470 & 0.02492(54) &$-$0.0004(11)  & 0.800(8)[50] & 4500 \\
      & 0.134630 & 0.01950(73) & 0.00175(52) & 0.814(8)[50] & 4500 \\
      & 0.134790 & 0.01474(63) &$-$0.00007(49) & 0.803(7)[50] & 4500 \\
      & 0.134950 & 0.00949(53) & 0.00038(65) & 0.804(8)[50] & 4500 \\
 1.35 & 0.133920 & 0.02369(45) & 0.00044(47) & 0.82(1)[50]  & 4500 \\
      & 0.134080 & 0.0190(10)  &$-$0.00040(73) & 0.803(7)[50] & 4500 \\
      & 0.134240 & 0.01426(70) & 0.00011(93) & 0.82(1)[50]  & 4500 \\
      & 0.134400 & 0.00907(51) &$-$0.00044(57) & 0.81(1)[50]  & 4500 \\
 \hline\hline\\[1ex]
 \multicolumn{6}{c}{$\beta= 2.20$, $L/a= 8$}\\[1ex]
 1.35 & 0.13868 & 0.01252(59) & 0.00182(70) & 0.828(3)[50] & 37300 \\
      & 0.13914 &$-$0.00310(93) & 0.00130(87) & 0.834(2)[50] & 35900 \\
 1.50 & 0.13654 & 0.00838(96) & 0.00071(72) & 0.834(2)[50] & 41100 \\
      & 0.13693 &$-$0.00433(75) & 0.00003(62) & 0.833(2)[50] & 39600 \\
 1.60 & 0.13500 & 0.01012(23) &$-$0.00060(39) & 0.839(2)[50] & 39900 \\
      & 0.13543 &$-$0.00483(38) &$-$0.00076(68) & 0.832(2)[50] & 38000 \\
 \hline\hline\\[1ex]
 \multicolumn{6}{c}{$\beta= 2.10$, $L/a= 8$}\\[1ex]
 1.38 & 0.14040 & 0.00598(95) & 0.0007(11)  & 0.811(2)[50] &145700 \\
      & 0.14092 &$-$0.01094(54) & 0.00188(56) & 0.810(2)[50] &167500 \\
 1.53 & 0.13741 & 0.02124(23) & 0.00062(65) & 0.829(2)[50] & 67800 \\
      & 0.13837 &$-$0.00872(53) & 0.00074(92) & 0.822(2)[50] & 58900 \\
 1.63 & 0.13599 & 0.01371(17) & 0.00030(58) & 0.830(2)[50] & 66400 \\
      & 0.13648 &$-$0.0018(11)  & 0.00020(57) & 0.828(2)[50] & 62900 \\
 1.73 & 0.13451 & 0.01152(32) &$-$0.00008(48) & 0.836(1)[50] &104400 \\
      & 0.13497 &$-$0.00317(40) &$-$0.0018(12)  & 0.833(2)[50] &137100 \\
 \hline\hline\\[1ex]
 \multicolumn{6}{c}{$\beta= 2.10$, $L/a= 6$}\\[1ex]
 1.2 & 0.14347 & 0.0077(22) & 0.0063(21)  & 0.864(3)[40] & 21600 \\
     & 0.14391 &$-$0.0025(21) & 0.0079(24)  & 0.865(4)[40] & 26000 \\
 1.4 & 0.13987 & 0.0125(18) & 0.0030(11)  & 0.868(4)[40] & 26000 \\
     & 0.14021 & 0.0043(14) & 0.0032(17)  & 0.864(3)[40] & 26000 \\
     & 0.14056 &$-$0.0090(23) & 0.0025(14)  & 0.862(4)[40] & 26000 \\
 1.6 & 0.13660 & 0.0073(13) & 0.0014(10)  & 0.869(7)[40] & 25200 \\
     & 0.13683 & 0.00065(93)&$-$0.00042(78) & 0.868(3)[40] & 26000 \\
     & 0.13725 &$-$0.0136(16) &$-$0.0002(11)  & 0.863(2)[40] & 25200 \\
 1.8 & 0.13335 & 0.00825(67)&$-$0.00438(74) & 0.867(4)[40] & 26000 \\
     & 0.13362 &$-$0.00169(79)&$-$0.00214(84) & 0.870(2)[40] & 26000 \\
     & 0.13389 &$-$0.0102(10) &$-$0.00278(77) & 0.867(4)[40] & 28000 \\
 2.0 & 0.13059 &$-$0.0001(14) &$-$0.00685(74) & 0.870(3)[40] & 21600 \\
     & 0.13090 &$-$0.0127(12) &$-$0.0071(10)  & 0.870(3)[40] & 21600 \\
 2.4 & 0.12500 & 0.0022(17) &$-$0.01283(75) & 0.873(3)[40] & 21600 \\
     & 0.12550 &$-$0.0148(31) &$-$0.0133(12)  & 0.865(3)[40] & 21600 \\
 \hline\hline\\[1ex]
 \multicolumn{6}{c}{$\beta= 2.00$, $L/a= 6$}\\[1ex]
 1.4 & 0.14279 & 0.0125(26) & 0.0068(17) & 0.906(4)[50] & 20000 \\
     & 0.14362 &$-$0.0135(21) & 0.0033(33) & 0.902(3)[50] & 25000 \\
 1.6 & 0.13901 & 0.0131(17) & 0.0034(21) & 0.914(2)[50] & 25000 \\
     & 0.13936 & 0.0035(16) & 0.00330(91)& 0.907(3)[50] & 25000 \\
     & 0.13971 &$-$0.0088(16) & 0.0009(26) & 0.906(4)[50] & 25000 \\
 1.8 & 0.13536 & 0.0113(15) & 0.0002(13) & 0.918(2)[50] & 25000 \\
     & 0.13588 &$-$0.0069(12) &$-$0.0013(22) & 0.909(4)[50] & 25000 \\
     & 0.13627 &$-$0.0198(45) & 0.0006(24) & 0.907(7)[50] & 25000 \\
 2.0 & 0.13192 & 0.0116(11) &$-$0.00490(92)& 0.918(3)[50] & 25000 \\
     & 0.13221 & 0.0015(18) &$-$0.0041(14) & 0.916(4)[50] & 25000 \\
     & 0.13250 &$-$0.0106(12) &$-$0.0033(11) & 0.915(2)[50] & 25000 \\
 2.2 & 0.12839 & 0.0207(11) &$-$0.0074(11) & 0.921(3)[50] & 20000 \\
     & 0.12902 &$-$0.0008(18) &$-$0.0078(12) & 0.919(2)[50] & 20000 \\
 2.6 & 0.12295 & 0.01399(86)&$-$0.01330(73)& 0.918(3)[50] & 21000 \\
     & 0.12353 &$-$0.0093(14) &$-$0.01204(69)& 0.921(3)[50] & 21000 \\
\hline\hline
\end{longtable*}
\begin{longtable*}{cccccc}
\caption[]{Same as Table~\ref{tab:Nf3MandDM} for quenched QCD.}
\endfirsthead
\caption[]{(\textit{Continued})}
\label{tab:Nf0MandDM}
\endhead
\endlastfoot
\endfoot
\hline\hline
 $\csw$ & $\kappa$ & $aM$  & $a\DM$ &$\PaccHMC[\NMD]$& $\Ntraj$
 \\\hline\\[1ex]
 \multicolumn{6}{c}{$\beta= 24.00$, $L/a= 8$}\\[1ex]
 1.00 & 0.12567 & 0.009136(59) & 0.000714(97) & 0.67(1)[128] & 3100 \\
      & 0.12584 & 0.003733(51) & 0.000545(69) & 0.67(1)[128] & 3100 \\
      & 0.12600 &$-$0.00139(11)  & 0.000525(96) & 0.65(2)[128] & 3100 \\
      & 0.12617 &$-$0.006783(76) & 0.000567(79) & 0.66(1)[128] & 3100 \\
 1.03 & 0.12567 & 0.007822(63) & 0.000078(90) & 0.67(1)[128] & 3100 \\
      & 0.12584 & 0.002319(54) & 0.000054(70) & 0.67(1)[128] & 3100 \\
      & 0.12600 &$-$0.002756(68) & 0.000167(81) & 0.67(1)[128] & 3100 \\
      & 0.12617 &$-$0.008282(63) &$-$0.000028(77) & 0.67(1)[128] & 3100 \\
 1.06 & 0.12567 & 0.006427(46) &$-$0.000334(63) & 0.67(1)[128] & 3100 \\
      & 0.12584 & 0.000984(78) &$-$0.000317(76) & 0.68(1)[128] & 3100 \\
      & 0.12600 &$-$0.00401(10)  &$-$0.000293(75) & 0.65(2)[128] & 3100 \\
      & 0.12617 &$-$0.009544(91) &$-$0.000295(78) & 0.66(1)[128] & 3100 \\
 \hline\hline\\[1ex]
 \multicolumn{6}{c}{$\beta= 12.00$, $L/a= 8$}\\[1ex]
 1.00 & 0.12659 & 0.013236(96)& 0.00080(17) & 0.74(2)[100] & 1600 \\
      & 0.12676 & 0.007870(89)& 0.00086(11) & 0.73(3)[100] & 1600 \\
      & 0.12693 & 0.00246(14) & 0.00086(17) & 0.72(1)[100] & 1600 \\
      & 0.12709 &$-$0.00261(10) & 0.00107(15) & 0.73(2)[100] & 1600 \\
 1.05 & 0.12659 & 0.009444(92)& 0.00023(14) & 0.74(2)[100] & 1600 \\
      & 0.12676 & 0.004108(91)& 0.00012(12) & 0.72(2)[100] & 1600 \\
      & 0.12693 &$-$0.001424(88)&$-$0.00000(15) & 0.70(2)[100] & 1600 \\
      & 0.12709 &$-$0.006318(91)& 0.000214(98)& 0.72(2)[100] & 1600 \\
 1.10 & 0.12659 & 0.005635(81)&$-$0.00043(13) & 0.73(2)[100] & 1600 \\
      & 0.12676 & 0.00026(16) &$-$0.00048(11) & 0.74(1)[100] & 1600 \\
      & 0.12693 &$-$0.00507(15) &$-$0.00053(10) & 0.73(2)[100] & 1600 \\
      & 0.12709 &$-$0.01021(11) &$-$0.00069(16) & 0.75(2)[100] & 1600 \\
 \hline\hline\\[1ex]
 \multicolumn{6}{c}{$\beta= 8.85$, $L/a= 8$}\\[1ex]
 1.05 & 0.1261 & 0.048752(68) & 0.000670(83) & 0.70(1)[80] & 3500 \\
      & 0.1266 & 0.033381(77) & 0.000472(91) & 0.70(1)[80] & 3500 \\
      & 0.1271 & 0.017522(76) & 0.000515(90) & 0.69(1)[80] & 3500 \\
      & 0.1276 & 0.001942(99) & 0.00034(11)  & 0.69(1)[80] & 3500 \\
 1.07 & 0.1260 & 0.049789(63) & 0.000197(85) & 0.70(1)[80] & 3500 \\
      & 0.1265 & 0.034276(80) & 0.000152(88) & 0.68(1)[80] & 3500 \\
      & 0.1270 & 0.018797(76) & 0.00027(12)  & 0.70(1)[80] & 3500 \\
      & 0.1275 & 0.00324(11)  & 0.00019(12)  & 0.69(1)[80] & 3500 \\
 1.09 & 0.1259 & 0.051029(60) &$-$0.000008(97) & 0.69(1)[80] & 3500 \\
      & 0.1264 & 0.035509(77) &$-$0.000033(81) & 0.70(1)[80] & 3500 \\
      & 0.1269 & 0.019930(91) &$-$0.00006(12)  & 0.71(1)[80] & 3500 \\
      & 0.1274 & 0.00427(13)  & 0.00005(12)  & 0.70(1)[80] & 3500 \\
 \hline\hline\\[1ex]
 \multicolumn{6}{c}{$\beta= 5.00$, $L/a= 8$}\\[1ex]
 1.08 & 0.12954 & 0.01759(11) & 0.00080(14) & 0.762(10)[64]& 4800 \\
      & 0.12970 & 0.01254(13) & 0.00075(25) & 0.751(9)[64] & 3500 \\
      & 0.12979 & 0.00991(18) & 0.00057(17) & 0.73(1)[64]  & 3500 \\
      & 0.12986 & 0.00767(17) & 0.00100(16) & 0.754(9)[64] & 3500 \\
      & 0.12995 & 0.00480(21) & 0.00115(19) & 0.75(1)[64]  & 3500 \\
      & 0.13002 & 0.00280(14) & 0.00083(25) & 0.75(1)[64]  & 3500 \\
      & 0.13011 &$-$0.00014(16) & 0.00094(27) & 0.738(9)[64] & 3500 \\
      & 0.13027 &$-$0.00508(20) & 0.00069(27) & 0.76(1)[64]  & 3500 \\
 1.13 & 0.12951 & 0.01003(15) &$-$0.00002(18) & 0.75(2)[64]  & 3500 \\
      & 0.12954 & 0.009028(95)& 0.00041(12) & 0.747(9)[64] & 3500 \\
      & 0.12967 & 0.00506(12) & 0.00012(26) & 0.75(1)[64]  & 3500 \\
      & 0.12970 & 0.00413(13) &$-$0.00009(29) & 0.725(10)[64]& 3500 \\
      & 0.12983 & 0.00026(12) & 0.00048(18) & 0.74(1)[64]  & 3500 \\
      & 0.12986 &$-$0.00082(15) & 0.00006(25) & 0.75(2)[64]  & 3500 \\
      & 0.12999 &$-$0.00489(18) & 0.00038(23) & 0.733(10)[64]& 3500 \\
      & 0.13002 &$-$0.00576(11) & 0.00019(20) & 0.747(9)[64] & 3500 \\
 1.18 & 0.12924 & 0.00995(12) &$-$0.00052(19) & 0.74(1)[64]  & 3500 \\
      & 0.12940 & 0.00500(16) &$-$0.00023(25) & 0.75(2)[64]  & 3500 \\
      & 0.12954 & 0.00075(15) &$-$0.00012(21) & 0.742(10)[64]& 3500 \\
      & 0.12956 & 0.00010(14) &$-$0.00005(20) & 0.75(1)[64]  & 3500 \\
      & 0.12970 &$-$0.00429(23) &$-$0.00019(32) & 0.73(2)[64]  & 3500 \\
      & 0.12972 &$-$0.00515(16) &$-$0.00054(22) & 0.75(1)[64]  & 3500 \\
      & 0.12986 &$-$0.00933(13) &$-$0.00018(27) & 0.74(2)[64]  & 3500 \\
      & 0.13002 &$-$0.01466(12) &$-$0.00032(17) & 0.755(9)[64] & 3500 \\
 \hline\hline\\[1ex]
 \multicolumn{6}{c}{$\beta= 3.00$, $L/a= 8$}\\[1ex]
 1.20 & 0.13393 & 0.01524(24) & 0.00083(26) & 0.785(6)[50] & 8100 \\
      & 0.13410 & 0.01010(25) & 0.00144(32) & 0.789(6)[50] & 8100 \\
      & 0.13428 & 0.00461(25) & 0.00152(41) & 0.786(8)[50] & 8100 \\
      & 0.13440 & 0.00120(27) & 0.00132(59) & 0.778(6)[50] & 8100 \\
 1.28 & 0.13315 & 0.01562(28) & 0.00057(29) & 0.77(1)[50]  & 8100 \\
      & 0.13332 & 0.01053(18) & 0.00010(29) & 0.782(8)[50] & 8100 \\
      & 0.13349 & 0.00520(30) & 0.00122(38) & 0.782(7)[50] & 8100 \\
      & 0.13363 & 0.00110(23) & 0.00003(36) & 0.777(7)[50] & 8100 \\
 1.36 & 0.13239 & 0.01422(36) &$-$0.00045(27) & 0.777(7)[50] & 8100 \\
      & 0.13255 & 0.00941(29) &$-$0.00050(34) & 0.776(7)[50] & 8100 \\
      & 0.13272 & 0.00414(21) & 0.00008(33) & 0.777(8)[50] & 8100 \\
      & 0.13286 &$-$0.00033(36) &$-$0.00049(35) & 0.776(9)[50] & 8100 \\
 \hline\hline\\[1ex]
 \multicolumn{6}{c}{$\beta= 2.70$, $L/a= 8$}\\[1ex]
 1.2 & 0.13605 & 0.01302(55) & 0.00276(40) & 0.803(4)[50] & 15000 \\
     & 0.13642 & 0.00235(76) & 0.00270(50) & 0.803(4)[50] & 15000 \\
     & 0.13680 &$-$0.01013(71) & 0.0029(12)  & 0.797(4)[50] & 15000 \\
 1.3 & 0.13472 & 0.01845(90) & 0.00187(60) & 0.798(8)[50] & 15000 \\
     & 0.13526 & 0.00207(59) & 0.00158(62) & 0.797(4)[50] & 15000 \\
     & 0.13544 &$-$0.00305(17) & 0.00150(24) & 0.800(5)[50] & 15000 \\
 1.4 & 0.13356 & 0.01833(30) & 0.00015(50) & 0.805(4)[50] & 15000 \\
     & 0.13412 & 0.00159(30) & 0.00113(48) & 0.798(8)[50] & 15000 \\
     & 0.13428 &$-$0.00405(34) & 0.00047(52) & 0.799(7)[50] & 15000 \\
 1.5 & 0.13264 & 0.01093(29) &$-$0.00113(28) & 0.800(5)[50] & 15000 \\
     & 0.13281 & 0.00562(27) &$-$0.00107(37) & 0.805(4)[50] & 15000 \\
     & 0.13298 &$-$0.0009(12)  &$-$0.00076(43) & 0.799(7)[50] & 15000 \\
 1.6 & 0.13143 & 0.01254(23) &$-$0.0037(14)  & 0.797(4)[50] & 15000 \\
     & 0.13180 & 0.00022(23) &$-$0.00202(31) & 0.815(6)[50] & 15000 \\
 \hline\hline\\[1ex]
 \multicolumn{6}{c}{$\beta= 2.70$, $L/a= 6$}\\[1ex]
 1.2 & 0.13605 & 0.01009(38) & 0.00323(43) & 0.888(3)[50] & 15000 \\
     & 0.13642 &$-$0.00071(39) & 0.00324(48) & 0.888(3)[50] & 15000 \\
     & 0.13680 &$-$0.01069(42) & 0.00298(64) & 0.890(4)[50] & 15000 \\
 1.3 & 0.13472 & 0.01651(28) & 0.00106(46) & 0.890(3)[50] & 15000 \\
     & 0.13526 & 0.00090(34) & 0.00097(53) & 0.890(4)[50] & 15000 \\
     & 0.13544 &$-$0.00499(36) & 0.00137(44) & 0.889(6)[50] & 15000 \\
 1.4 & 0.13356 & 0.01695(39) &$-$0.00102(36) & 0.889(3)[50] & 15000 \\
     & 0.13412 & 0.00017(32) &$-$0.00088(53) & 0.890(3)[50] & 15000 \\
     & 0.13428 &$-$0.00464(39) &$-$0.00070(51) & 0.890(3)[50] & 15000 \\
 1.5 & 0.13264 & 0.00959(32) &$-$0.00261(41) & 0.889(6)[50] & 15000 \\
     & 0.13298 &$-$0.00068(35) &$-$0.00267(42) & 0.890(3)[50] & 15000 \\
 1.6 & 0.13143 & 0.01083(32) &$-$0.00504(54) & 0.890(4)[50] & 15000 \\
 \hline\hline\\[1ex]
 \multicolumn{6}{c}{$\beta= 2.60$, $L/a= 6$}\\[1ex]
 1.2 & 0.13698 & 0.00983(59) & 0.00469(88) & 0.896(3)[50] & 15000 \\
     & 0.13730 & 0.00064(65) & 0.00482(99) & 0.896(3)[50] & 15000 \\
     & 0.13749 &$-$0.00550(60) & 0.0049(11)  & 0.893(7)[50] & 15000 \\
 1.3 & 0.13574 & 0.01057(49) & 0.00281(79) & 0.893(7)[50] & 15000 \\
     & 0.13616 &$-$0.00198(52) & 0.00170(81) & 0.894(3)[50] & 15000 \\
 1.4 & 0.13463 & 0.00750(57) &$-$0.00051(53) & 0.894(3)[50] & 15000 \\
     & 0.13494 &$-$0.00187(34) & 0.00068(58) & 0.894(3)[50] & 15000 \\
 1.5 & 0.13331 & 0.01048(54) &$-$0.00249(50) & 0.891(3)[50] & 15000 \\
     & 0.13367 &$-$0.00005(38) &$-$0.00176(56) & 0.893(3)[50] & 15000 \\
 1.6 & 0.13215 & 0.00774(45) &$-$0.00342(50) & 0.894(3)[50] & 15000 \\
 1.8 & 0.12953 & 0.01242(62) &$-$0.00778(50) & 0.893(3)[50] & 15000 \\
\hline\hline
\end{longtable*}
\clearpage
\begin{table}
\centering
\caption{Numerical values of $\csw(\gbare,L/a)$ and
 $\kappa(\gbare,L/a)$ satisfying Eq.~(\ref{eq:setup:ImpCnd-new}) in
 three-flavor QCD.}
\label{tab:Nf3NPTKCSW}
\begin{ruledtabular}
 \begin{tabular}{ccccc}
$\beta$& $L/a$
  & function
  & $\csw(\gbare,L/a)$ 
  & $\kappa(\gbare,L/a)$ \\\hline
 12.00 & 8 & linear 
       & 1.0546(25) & 0.1268421(61) \\
  8.85 & 8 & linear 
       & 1.0761(32) & 0.127513(10) \\
  5.00 & 8 & linear 
       & 1.1311(48) & 0.129641(26) \\
  3.00 & 8 & linear 
       & 1.254(15)  & 0.13318(14) \\
  2.60 & 8 & linear 
       & 1.359(13)  & 0.13423(14) \\
  2.40 & 8 & linear 
       & 1.384(23)  & 0.13545(29) \\
  2.20 & 8 & linear 
       & 1.508(29)  & 0.13587(39) \\
  2.10 & 8 & linear 
       & 1.649(58)  & 0.13521(85) \\
  2.00 & 8 & quadratic 
       & 1.670(56)  & 0.13639(89) \\
  2.00 & 6 & quadratic 
       & 1.632(45)  & 0.13696(77) \\
  1.90 & 6 & quadratic 
       & 1.739(53)  & 0.13741(98) \\
 \end{tabular}
 \end{ruledtabular}
\end{table}
\begin{table}
\centering
\caption{Same as Table~\ref{tab:Nf3NPTKCSW}, but for two-flavor QCD.}
\label{tab:Nf2NPTKCSW}
 \begin{ruledtabular}
 \begin{tabular}{ccccc}
$\beta$ & $L/a$ & function 
        & $\csw(\gbare,L/a)$ & $\kappa(\gbare,L/a)$  \\\hline
 12.00 & 8 & linear 
       & 1.0558(27)& 0.1268509(66)\\
  8.85 & 8 & linear 
       & 1.0818(85)& 0.127519(27) \\
  5.00 & 8 & linear 
       & 1.1334(62)& 0.129686(34) \\
  3.00 & 8 & linear 
       & 1.276(20) & 0.13320(19)  \\
  2.60 & 8 & linear 
       & 1.327(49) & 0.13496(55)  \\
  2.20 & 8 & linear 
       & 1.519(32) & 0.13649(48)  \\
  2.10 & 8 & linear 
       & 1.672(65) & 0.1358(10)   \\
  2.10 & 6 & linear 
       & 1.598(19) & 0.13689(31)  \\
  2.00 & 6 & linear 
       & 1.777(27) & 0.13612(47) \\
 \end{tabular}
 \end{ruledtabular}
\end{table}
\begin{table}
\centering
\caption{Same as Table~\ref{tab:Nf3NPTKCSW}, but for quenched QCD.}
\label{tab:Nf0NPTKCSW}
 \begin{ruledtabular}
 \begin{tabular}{ccccc}
$\beta$ & $L/a$ & function 
        & $\csw(\gbare,L/a)$ & $\kappa(\gbare,L/a)$ \\\hline
24.00 & 8 & linear
      & 1.0375(16) & 0.1259026(23) \\
12.00 & 8 & linear
      & 1.0627(27) & 0.1268574(65) \\
 8.85 & 8 & linear
      & 1.0829(47) & 0.127565(15) \\
 5.00 & 8 & linear
      & 1.1540(42) & 0.129701(23) \\
 3.00 & 8 & linear
      & 1.338(20)  & 0.13308(19) \\
 2.70 & 8 & linear
      & 1.429(10)  & 0.13380(12) \\
 2.70 & 6 & linear
      & 1.3608(79) & 0.134554(92) \\
 2.60 & 6 & linear
      & 1.414(14)  & 0.13470(17) \\
 \end{tabular}
 \end{ruledtabular}
\end{table}
\begin{table}
\centering
\caption{Final results for $\csw(\gbare,L^{*}/a)$ and $\kappa(\gbare,L^{*}/a)$ for fixed physical size $L^*$ for three-flavor QCD.}
\label{tab:Nf3NPTKCSW_b19_l6}
 \begin{ruledtabular}
 \begin{tabular}{cclcl}
$\beta$ & $L/a$ & $L^{*}/a$
        & $\csw(\gbare,L^{*}/a)$ & $\kappa_c(\gbare,L^{*}/a)$
 \\\hline
 12.00 & 8 & 7.508095 $\times 10^{ 6}$ & 1.0601(25) & 0.1269060(61) \\
  8.85 & 8 & 8.462365 $\times 10^{ 4}$ & 1.0903(32) & 0.127592(10) \\
  5.00 & 8 & 3.807760 $\times 10^{ 2}$ & 1.1825(48) & 0.129764(26) \\
  3.00 & 8 & 2.502040 $\times 10^{ 1}$ & 1.368(15)  & 0.13325(14) \\
  2.60 & 8 & 1.475172 $\times 10^{ 1}$ & 1.467(13)  & 0.13424(14) \\
  2.40 & 8 & 1.136512 $\times 10^{ 1}$ & 1.471(23)  & 0.13544(29) \\
  2.20 & 8 & 8.780129 & 1.542(29) & 0.13586(39) \\
  2.10 & 8 & 7.726477 & 1.633(58) & 0.13521(85) \\
  2.00 & 8 & 6.805369 & 1.576(56) & 0.13642(89) \\
  2.00 & 6 & 6.805369 & 1.742(45) & 0.13691(77) \\
  2.00 & 6.805369 & 6.805369 & 1.650(51) & 0.13669(83) \\
  1.90 & 6 & 6 & 1.739(53) & 0.13741(98) \\
 \end{tabular}
 \end{ruledtabular}
\end{table}
\begin{table}
\centering
\caption{Same as Table~\ref{tab:Nf3NPTKCSW_b19_l6}, but for two-flavor QCD.}
\label{tab:Nf2NPTKCSW_b20_l6}
 \begin{ruledtabular}
 \begin{tabular}{cclcl}
$\beta$ & $L/a$ & $L^{*}/a$
        & $\csw(\gbare,L^{*}/a)$ & $\kappa(\gbare,L^{*}/a)$ 
  \\\hline
 12.00 & 8 & 2.350129 $\times 10^{ 6}$ & 1.0583(27) & 0.1269100(66) \\
  8.85 & 8 & 3.656026 $\times 10^{ 4}$ & 1.0919(85) & 0.127592(27) \\
  5.00 & 8 & 2.446546 $\times 10^{ 2}$ & 1.1797(62) & 0.129794(34) \\
  3.00 & 8 & 1.982120 $\times 10^{ 1}$ & 1.371(20)  & 0.13323(19) \\
  2.60 & 8 & 1.219378 $\times 10^{ 1}$ & 1.405(49)  & 0.13495(55) \\
  2.20 & 8 & 7.575548 & 1.498(32)  & 0.13651(48) \\
  2.10 & 8 & 6.738767 & 1.590(65)  & 0.1358(10) \\
  2.10 & 6 & 6.738767 & 1.682(19)  & 0.13680(31) \\
  2.10 & 6.738767 & 6.738767 & 1.631(39)  & 0.13638(63) \\
  2.00 & 6 & 6 & 1.777(27)  & 0.13612(47) \\
 \end{tabular}
 \end{ruledtabular}
\end{table}
\begin{table}[h]
\centering
\caption{Same as Table~\ref{tab:Nf3NPTKCSW_b19_l6}, but for quenched QCD.}
\label{tab:Nf0NPTKCSW_b26_l6}
 \begin{ruledtabular}
 \begin{tabular}{cclcl}
$\beta$ & $L/a$ & $L^{*}/a$
        & $\csw(\gbare,L^{*}/a)$ & $\kappa(\gbare,L^{*}/a)$
  \\\hline
24.00 & 8 & 3.088560 $\times 10^{11}$ & 1.0264(16) & 0.1259370(23) \\
12.00 & 8 & 2.409888 $\times 10^{ 5}$ & 1.0590(27) & 0.1269080(65) \\
 8.85 & 8 & 6.326167 $\times 10^{ 3}$ & 1.0852(47) & 0.127627(15) \\
 5.00 & 8 & 8.042260 $\times 10^{ 1}$ & 1.1921(42) & 0.129763(23) \\
 3.00 & 8 & 9.115448                  & 1.359(20)  & 0.13305(19) \\
 2.70 & 8 & 6.655769                  & 1.381(10)  & 0.13389(12) \\
 2.70 & 6 & 6.655769                  & 1.4006(79) & 0.134470(92) \\
 2.70 & 6.655769 & 6.655769           & 1.388(9)   & 0.13426(10) \\
 2.60 & 6 & 6                         & 1.414(14)  & 0.13470(17) \\
 \end{tabular}
 \end{ruledtabular}
\end{table}
\begin{table}
\centering
\caption{$L^*/a$, $\delta\csw$ and $\delta\kappa_c$ with the three-loop
 $\beta$ function Eq.~(\ref{eq:scaling-3loop}).}
\label{tab:sys error 1}
  \begin{ruledtabular}
  \begin{tabular}{rclrr}
$\beta$ & $L/a$ & $L^{*}/a$
        & $\delta     \csw(\gbare,L/a;L^{*}/a)$ 
        & $\delta \kappa_c(\gbare,L/a;L^{*}/a)$\\\hline
 12.00 & 8 & 1.688064 $\times 10^{ 7}$
       &   5.509123$\times10^{-3}$ &   6.348932$\times10^{-5}$\\
  8.85 & 8 & 1.755852 $\times 10^{ 5}$
       &   1.412367$\times10^{-2}$ &   7.951364$\times10^{-5}$\\
  5.00 & 8 & 6.442878 $\times 10^{ 2}$
       &   4.794420$\times10^{-2}$ &   1.251814$\times10^{-4}$\\
  3.00 & 8 & 3.300322 $\times 10^{ 1}$
       &   1.179115$\times10^{-1}$ &   9.547393$\times10^{-5}$\\
  2.60 & 8 & 1.793750 $\times 10^{ 1}$
       &   1.198918$\times10^{-1}$ &   3.523283$\times10^{-5}$\\
  2.40 & 8 & 1.317591 $\times 10^{ 1}$
       &   1.058685$\times10^{-1}$ &   1.836828$\times10^{-6}$\\
  2.20 & 8 & 9.648020
       &   6.105866$\times10^{-2}$ &$-$1.406025$\times10^{-5}$\\
  2.10 & 8 & 8.244442
       &   1.257961$\times10^{-2}$ &$-$4.125600$\times10^{-6}$\\
  2.00 & 8 & 7.037491
       &$-$7.046997$\times10^{-2}$ &   2.824179$\times10^{-5}$\\
  2.00 & 6 & 7.037491
       &   1.335089$\times10^{-1}$ &$-$6.098453$\times10^{-5}$\\
  1.90 & 6 & 6                         &   0 &  0\\
  \end{tabular}
  \end{ruledtabular}
\end{table}

\clearpage
\begin{widetext}
\newcommand{\figscale}{0.45}
\begin{figure*}
\centering
 \caption{$aM$ dependence of $a\DM$ in three-flavor QCD.}
 \label{fig:MandDM_3f}
\begin{tabular}{cc}
\includegraphics[scale=\figscale,clip=]{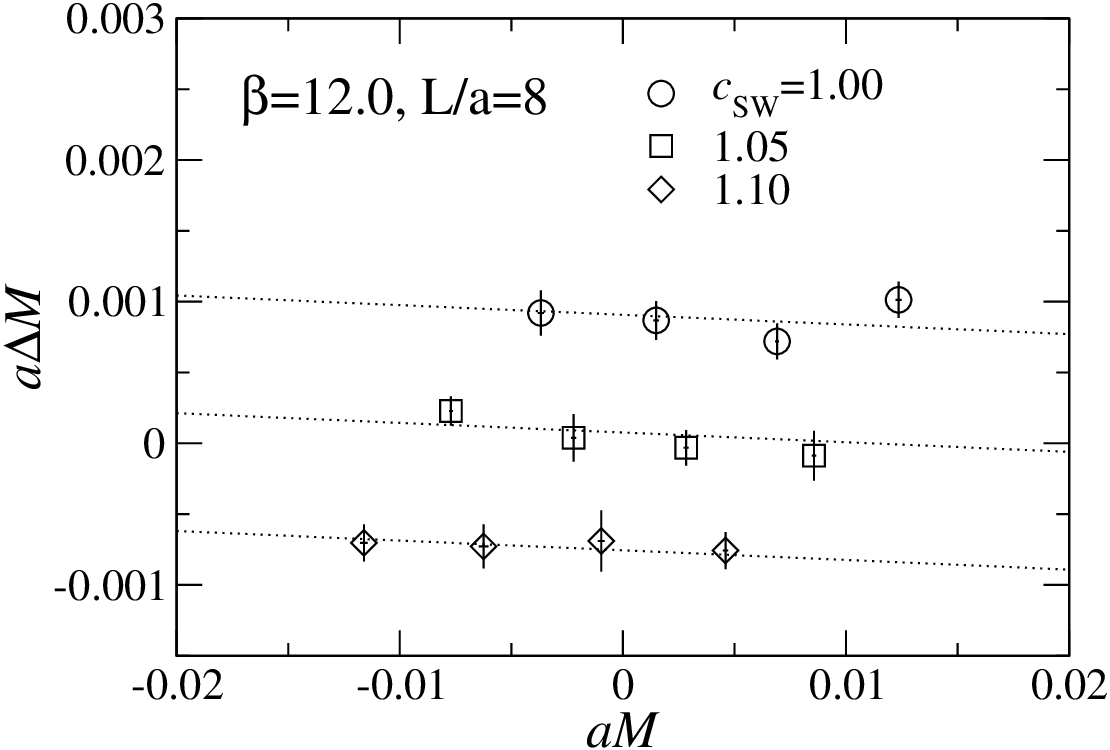} &
\includegraphics[scale=\figscale,clip=]{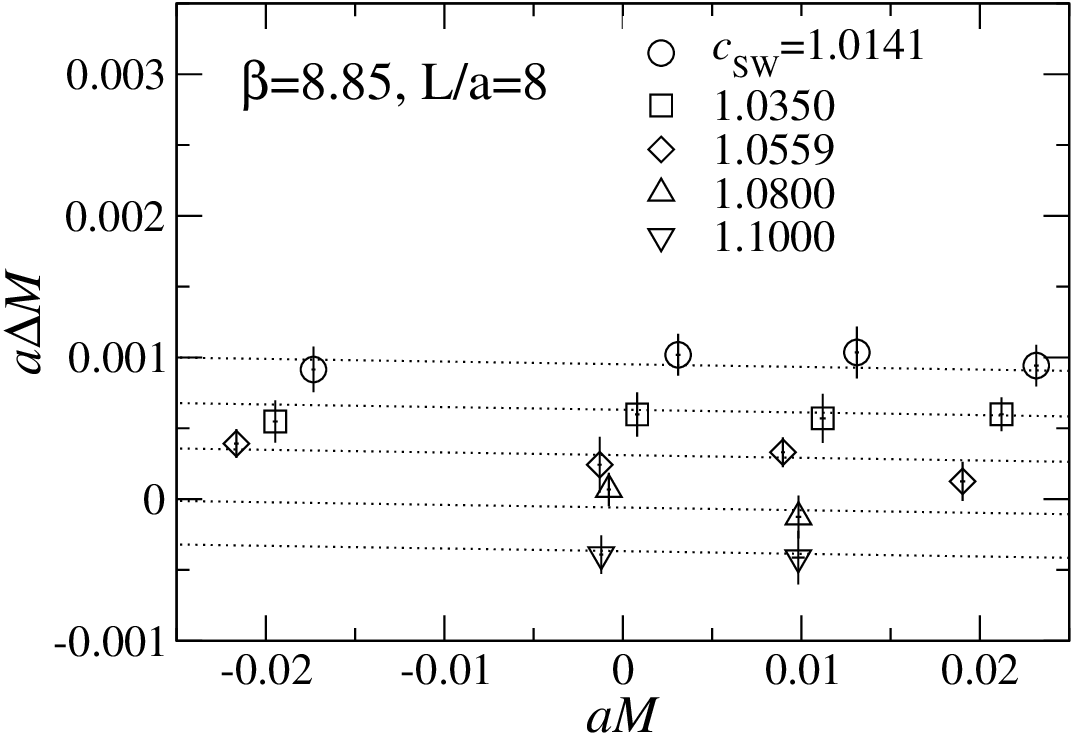} \\
\includegraphics[scale=\figscale,clip=]{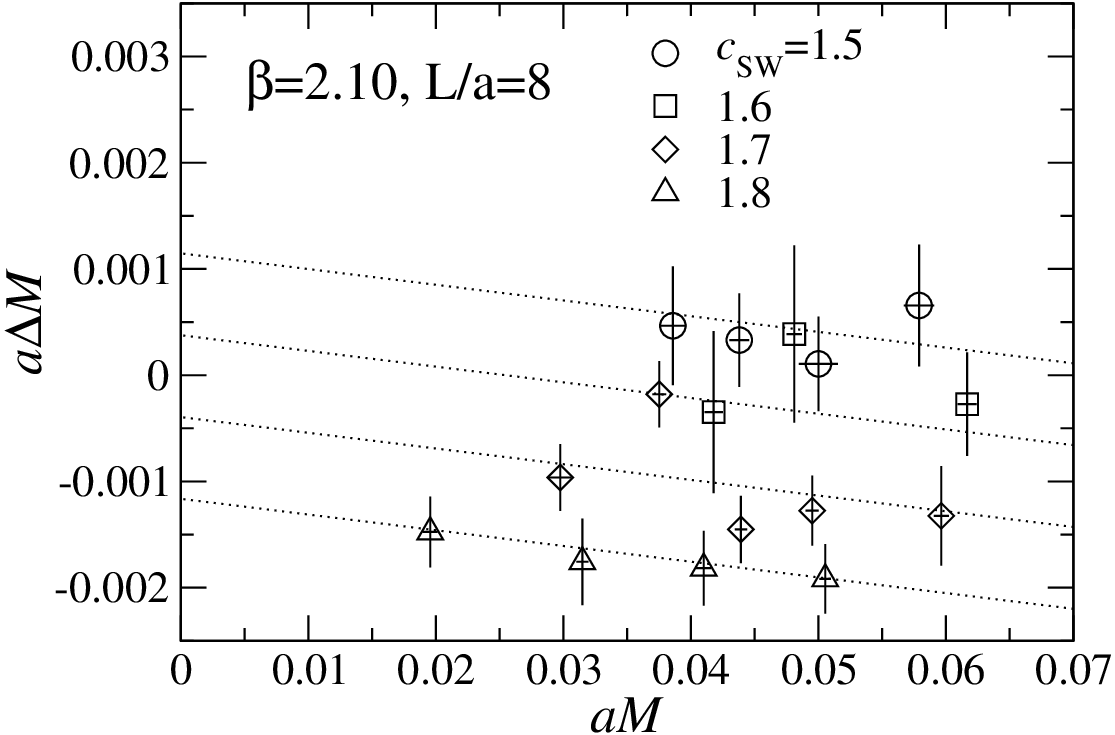} &
\includegraphics[scale=\figscale,clip=]{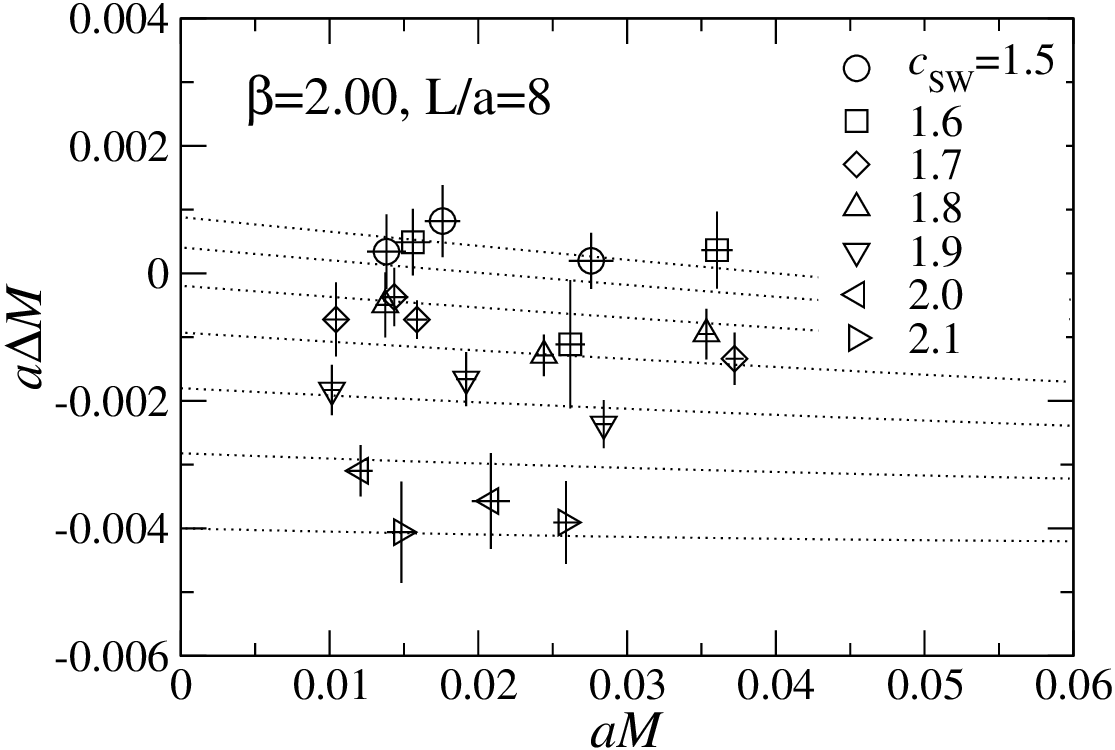} \\
\includegraphics[scale=\figscale,clip=]{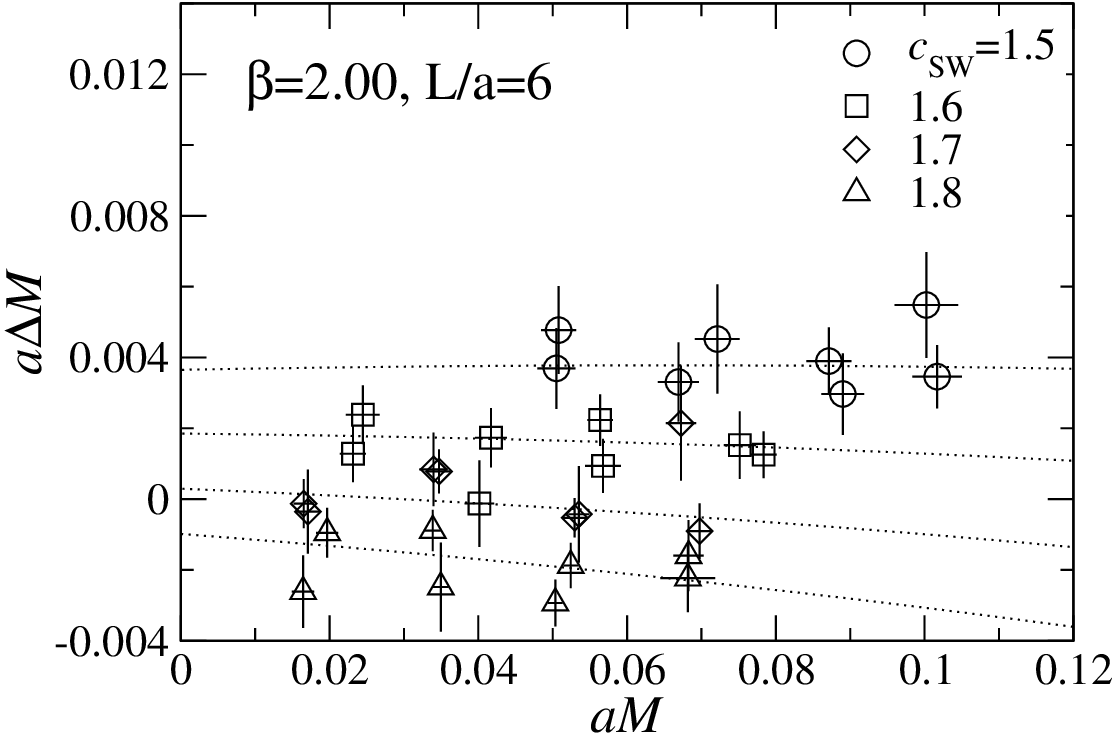} &
\includegraphics[scale=\figscale,clip=]{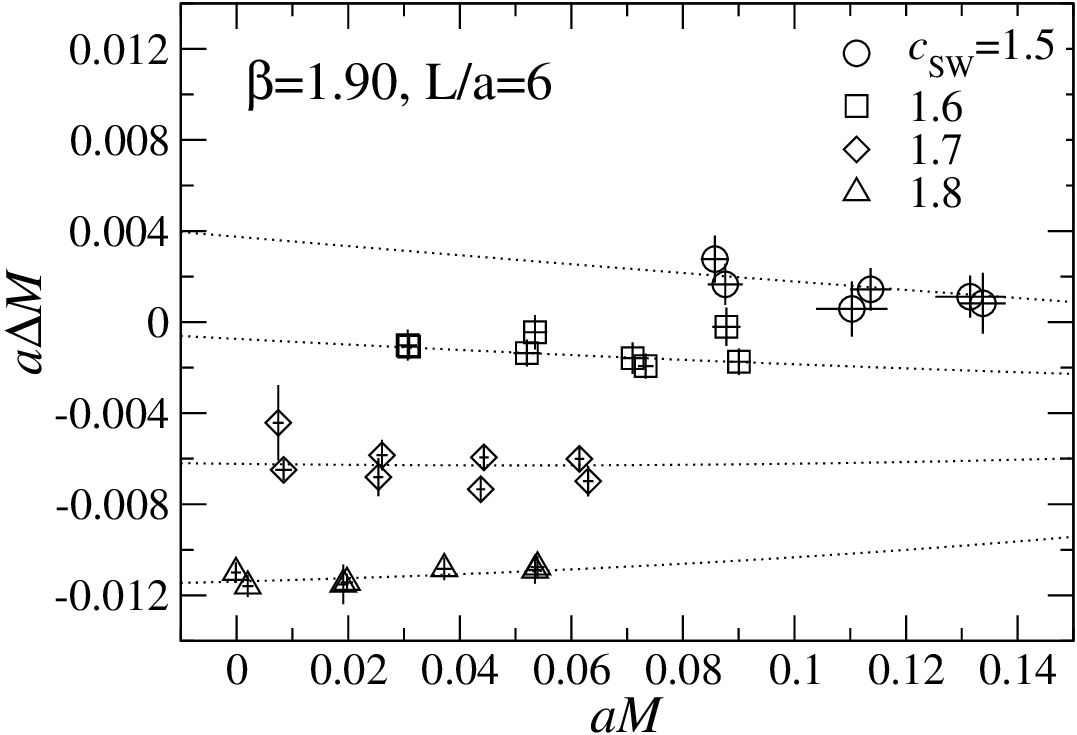} \\
\end{tabular}
\end{figure*}
\begin{figure*}
 \centering
 \caption{Same as Fig.~\ref{fig:MandDM_3f}, but in two-flavor QCD.}
 \label{fig:MandDM_2f}
\begin{tabular}{cc}
\includegraphics[scale=\figscale,clip=]{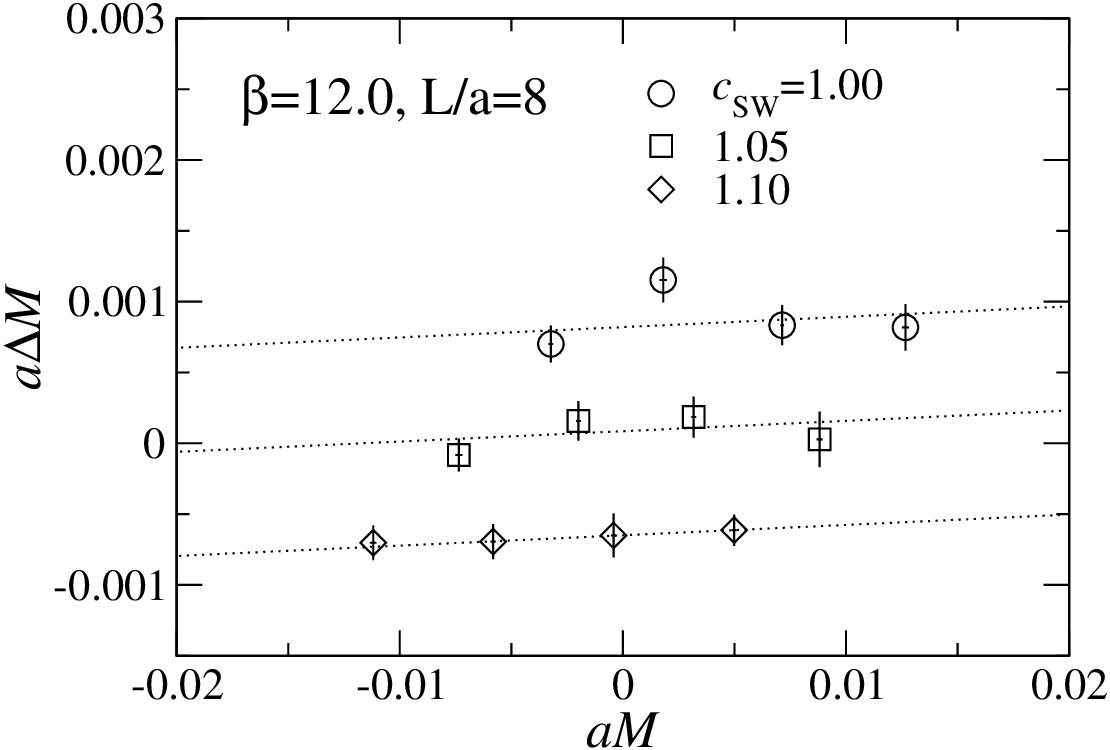} &
\includegraphics[scale=\figscale,clip=]{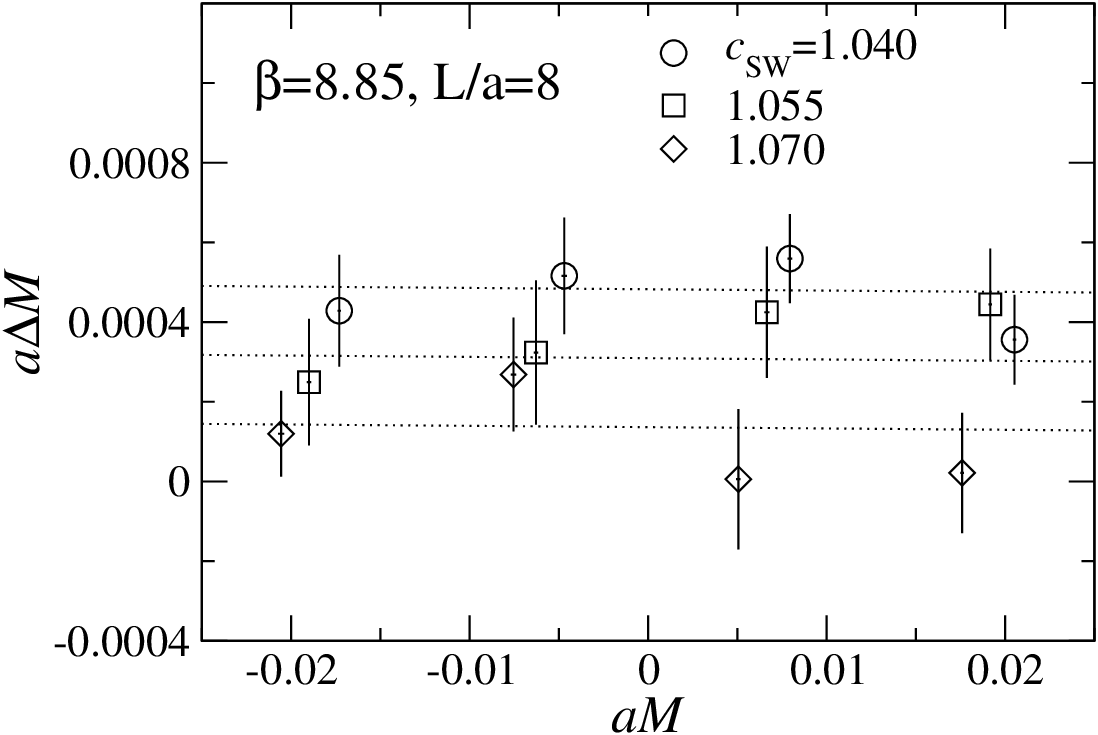} \\
\includegraphics[scale=\figscale,clip=]{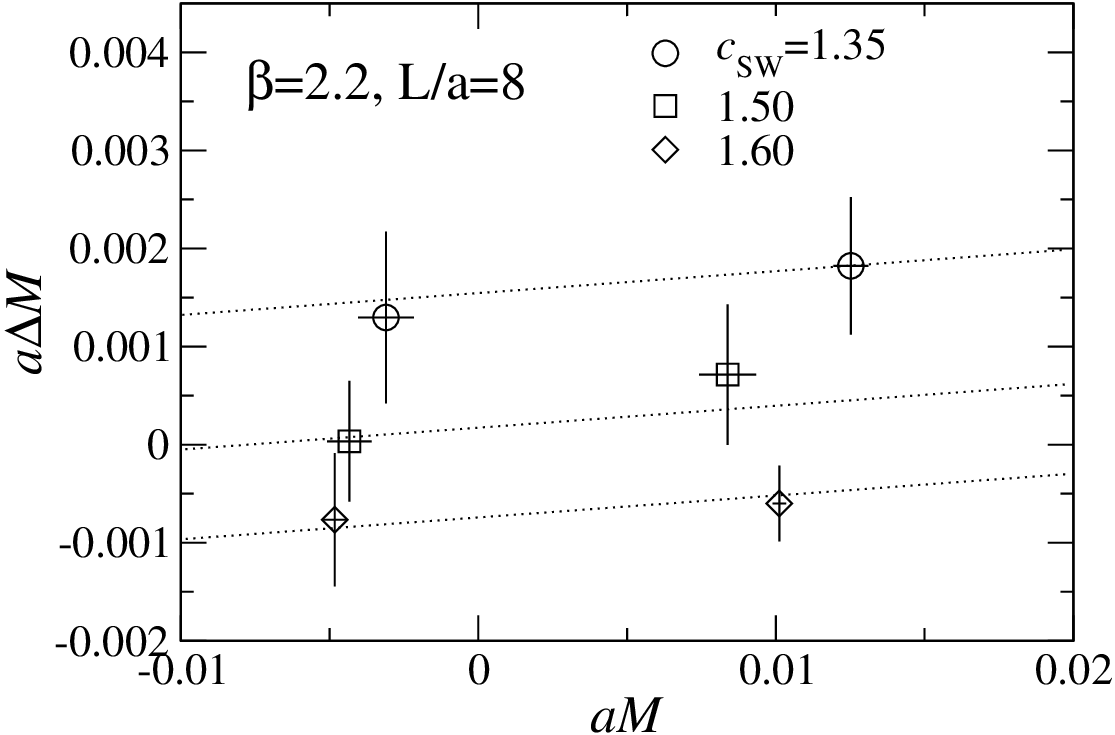} &
\includegraphics[scale=\figscale,clip=]{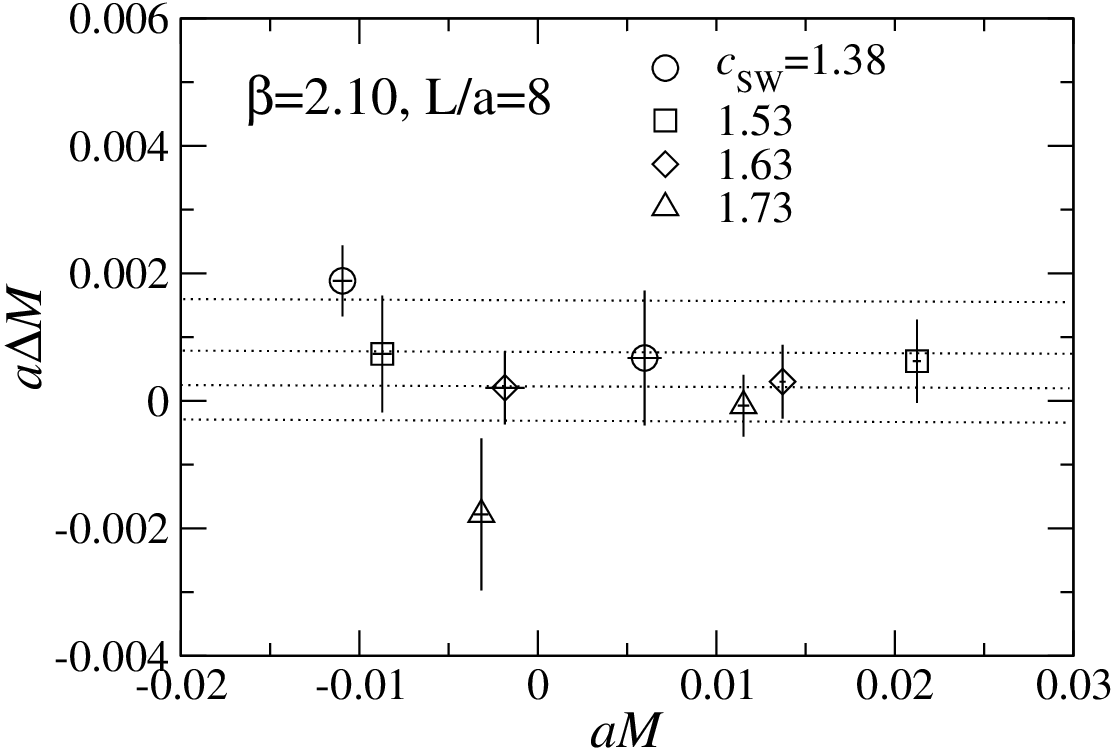} \\
\includegraphics[scale=\figscale,clip=]{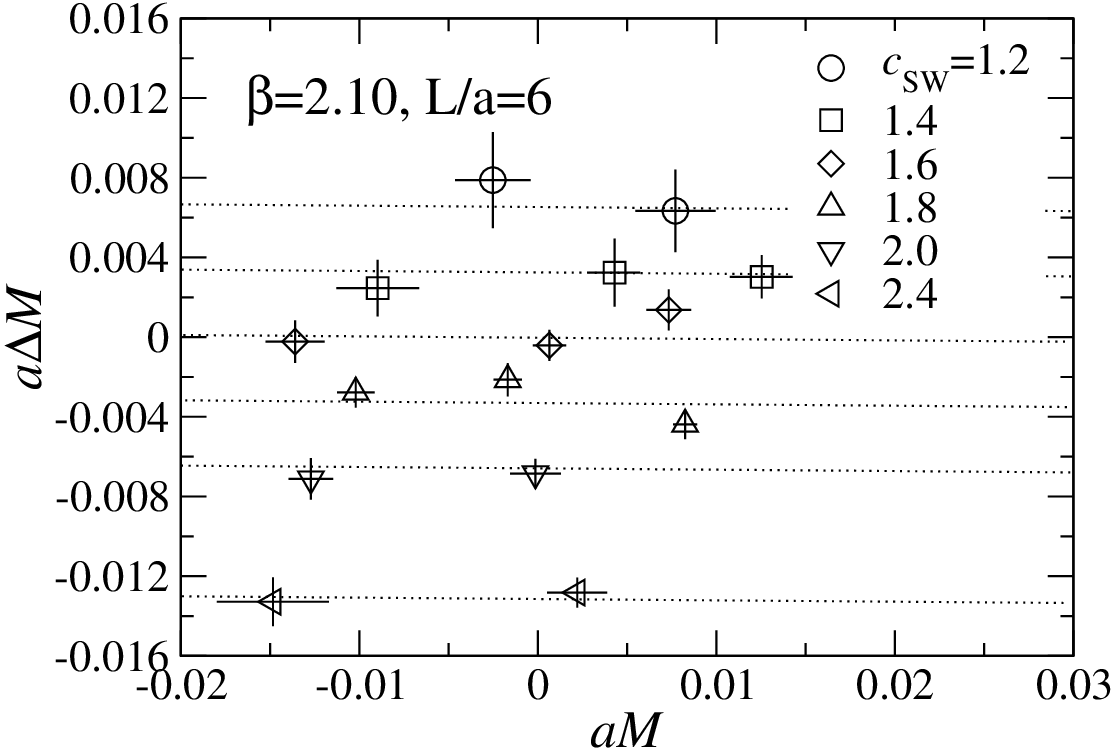} &
\includegraphics[scale=\figscale,clip=]{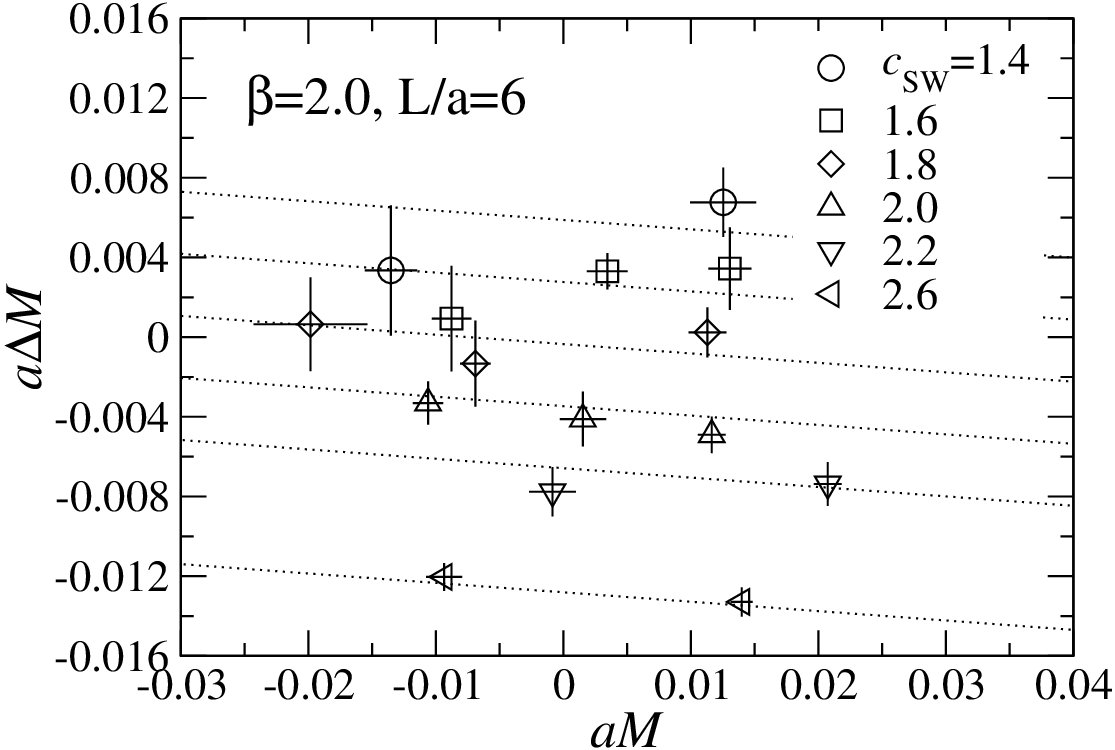} \\
\end{tabular}
\end{figure*}
\begin{figure*}
\centering
 \caption{Same as Fig.~\ref{fig:MandDM_3f}, but in quenched QCD.}
 \label{fig:MandDM_0f}
\begin{tabular}{cc}
\includegraphics[scale=\figscale,clip=]{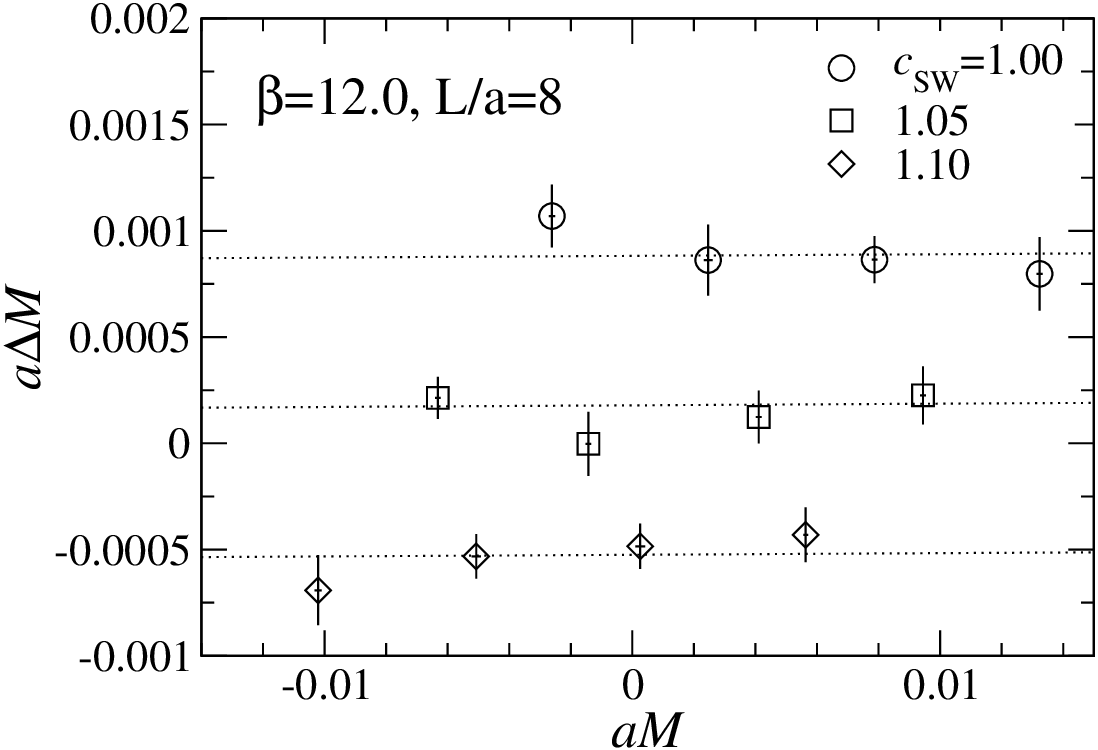} &
\includegraphics[scale=\figscale,clip=]{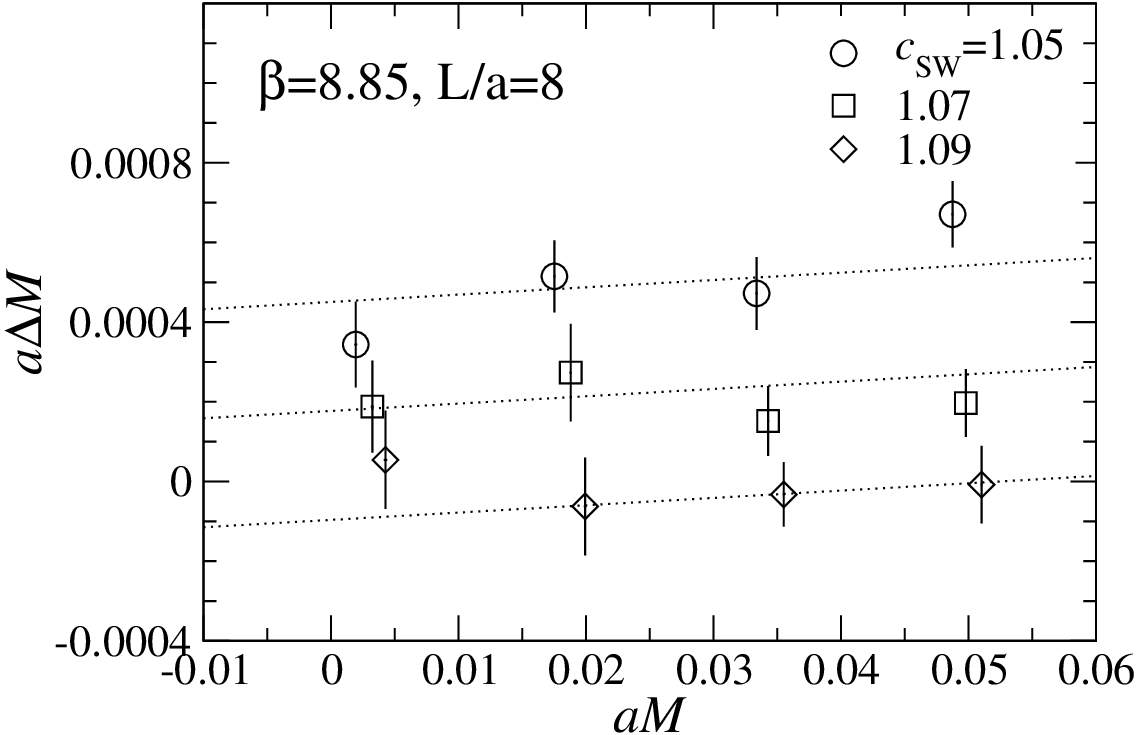} \\
\includegraphics[scale=\figscale,clip=]{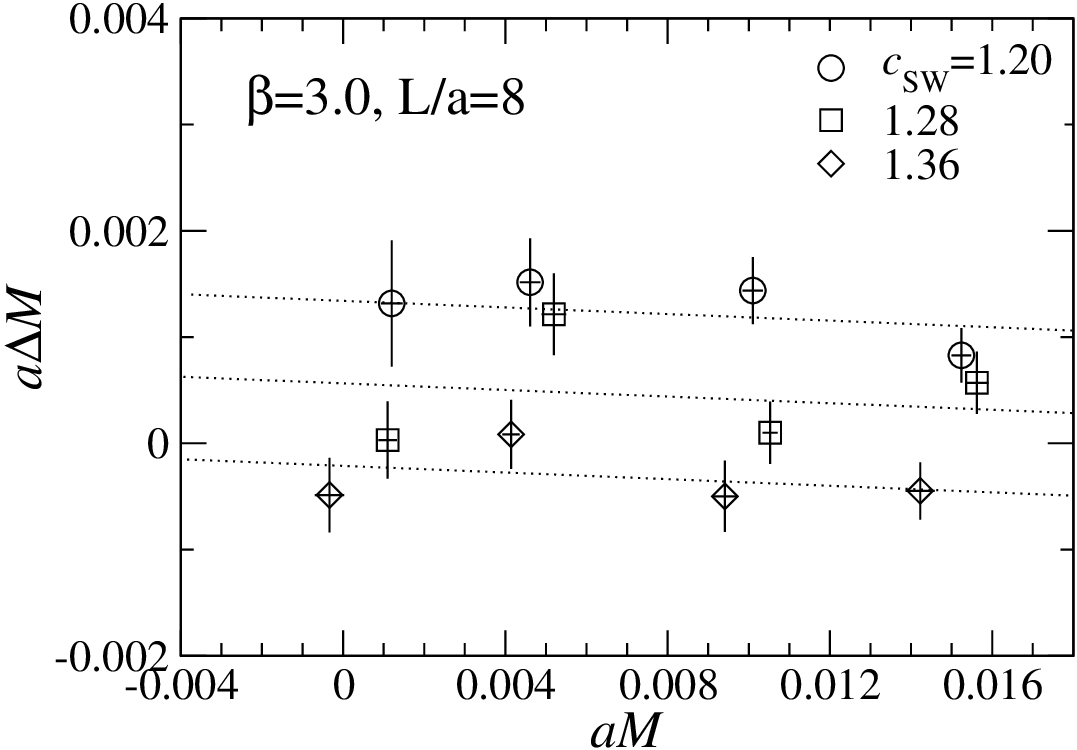} &
\includegraphics[scale=\figscale,clip=]{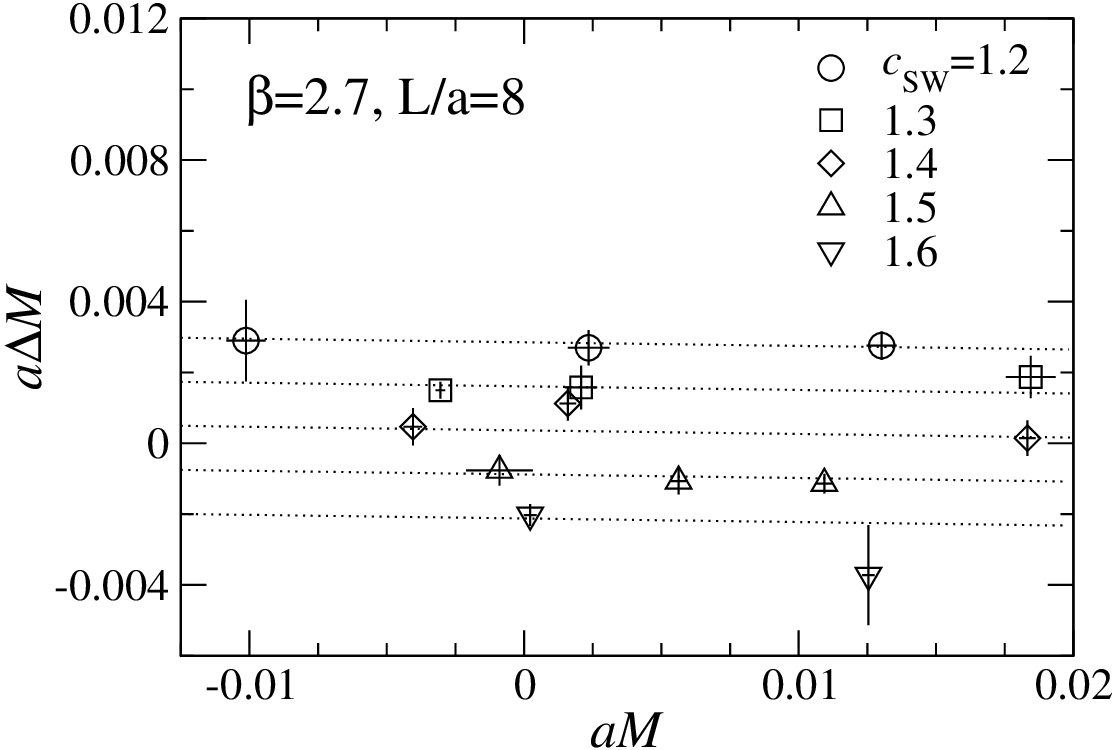} \\
\includegraphics[scale=\figscale,clip=]{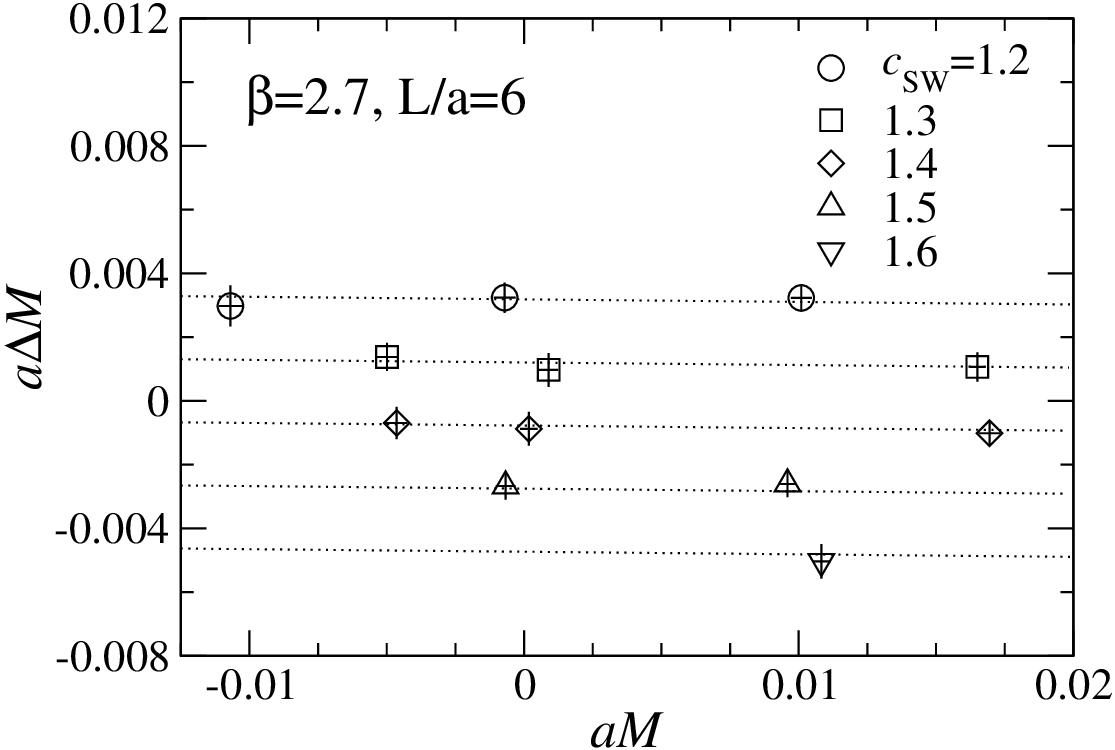} &
\includegraphics[scale=\figscale,clip=]{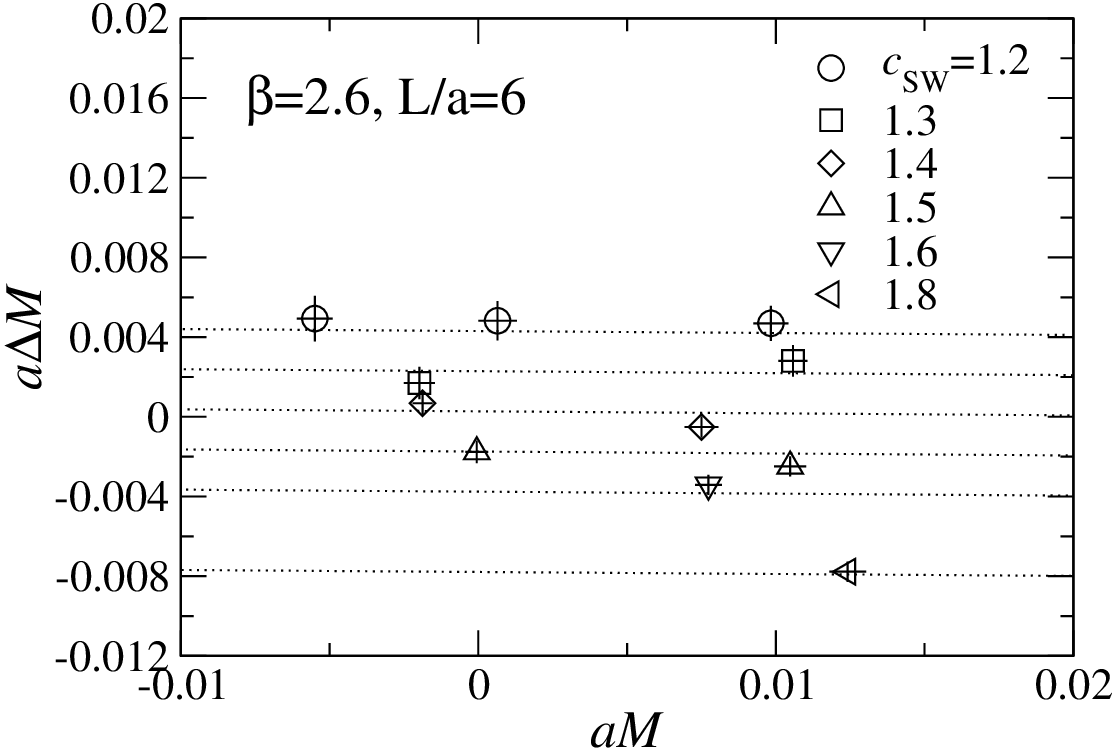} \\
\end{tabular}
\end{figure*}
\renewcommand{\figscale}{0.6}
\begin{figure}
\centering
\caption{$\gbare$ dependence of $\cswnp(\gbare,L^*/a)$ in $N_f$=3, 2,
 and 0 flavor QCD from top to bottom.
 Filled symbols are used for curve fitting.}
\includegraphics[scale=\figscale,clip=]{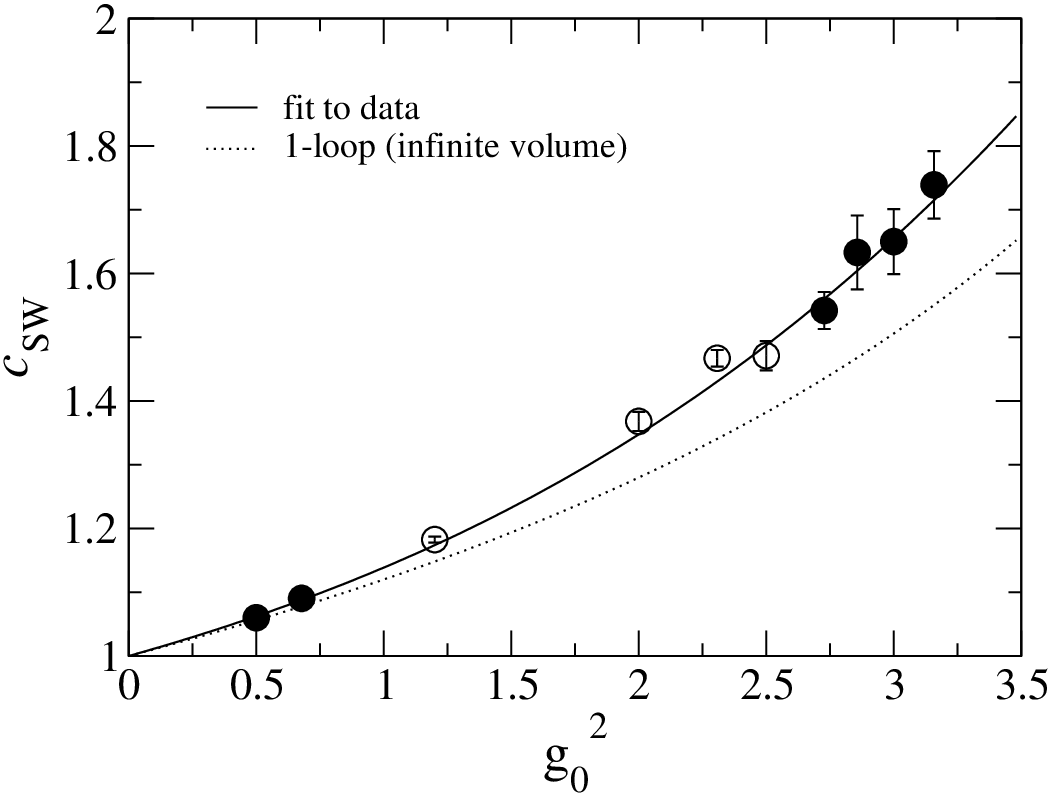}\\
\includegraphics[scale=\figscale,clip=]{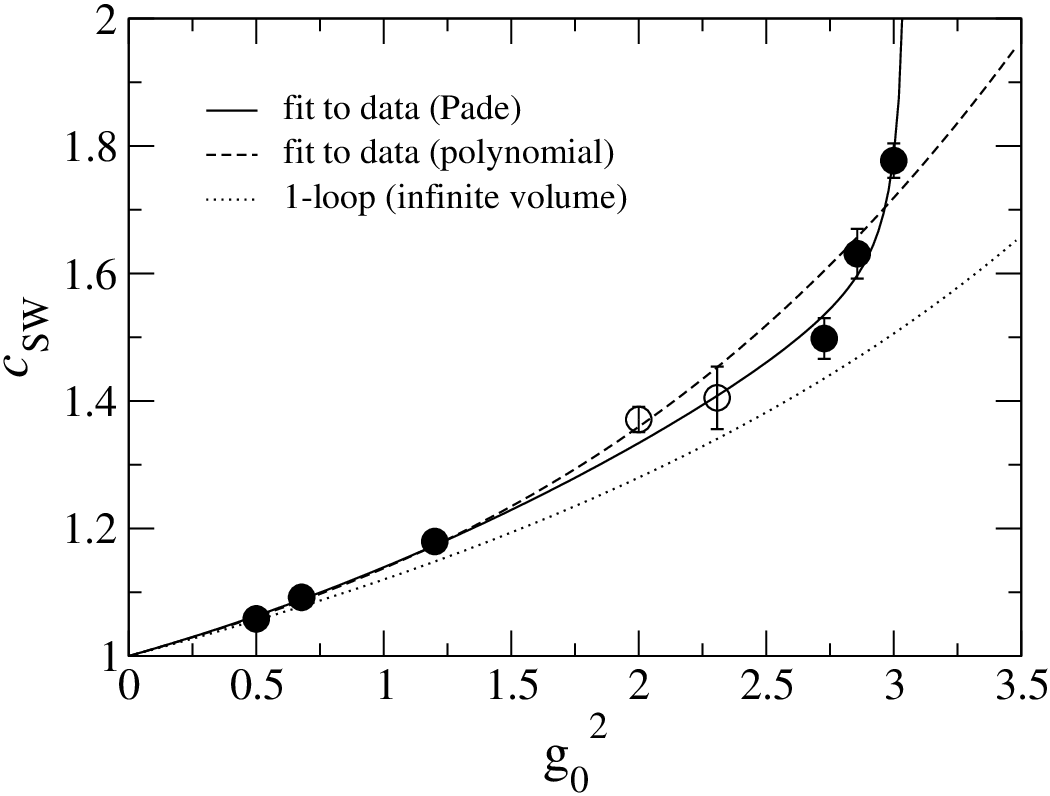}\\
\includegraphics[scale=\figscale,clip=]{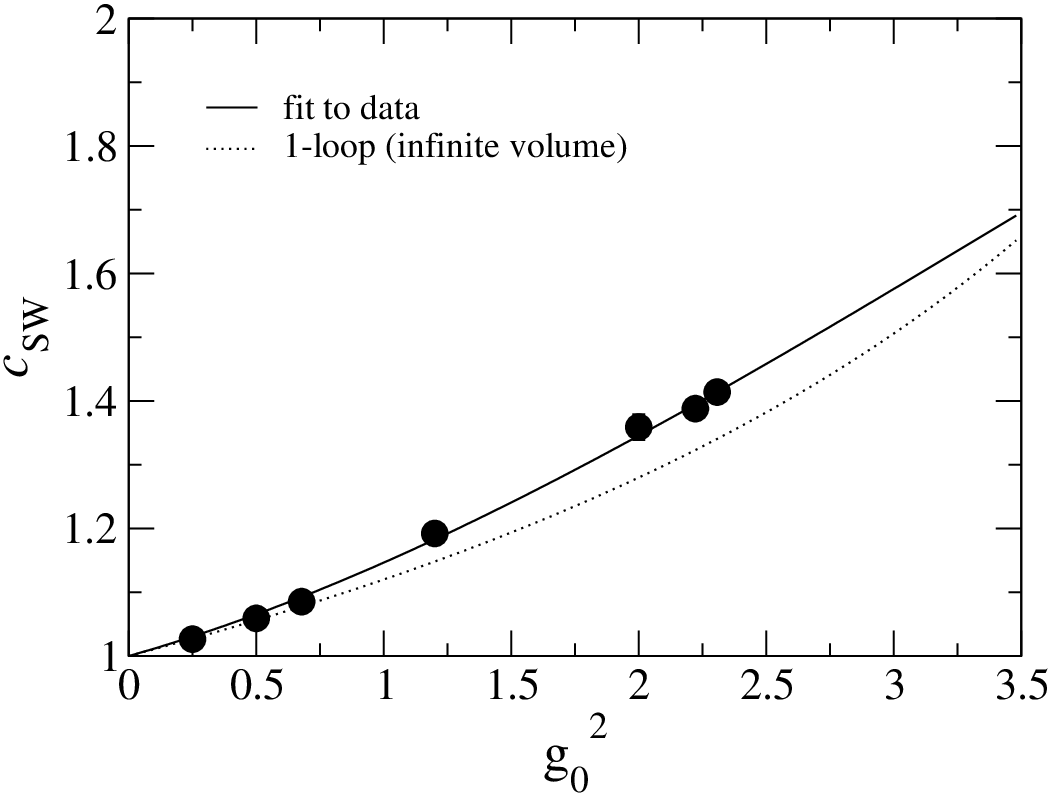}
\label{fig:Nf320Csw}
\end{figure}
\begin{figure}
\centering
\caption{$\gbare$ dependence of $\kappa_c(\gbare,L^*/a)$ in $N_f$=3, 2,
 and 0 flavor QCD from top to bottom.
}
\includegraphics[scale=\figscale,clip=]{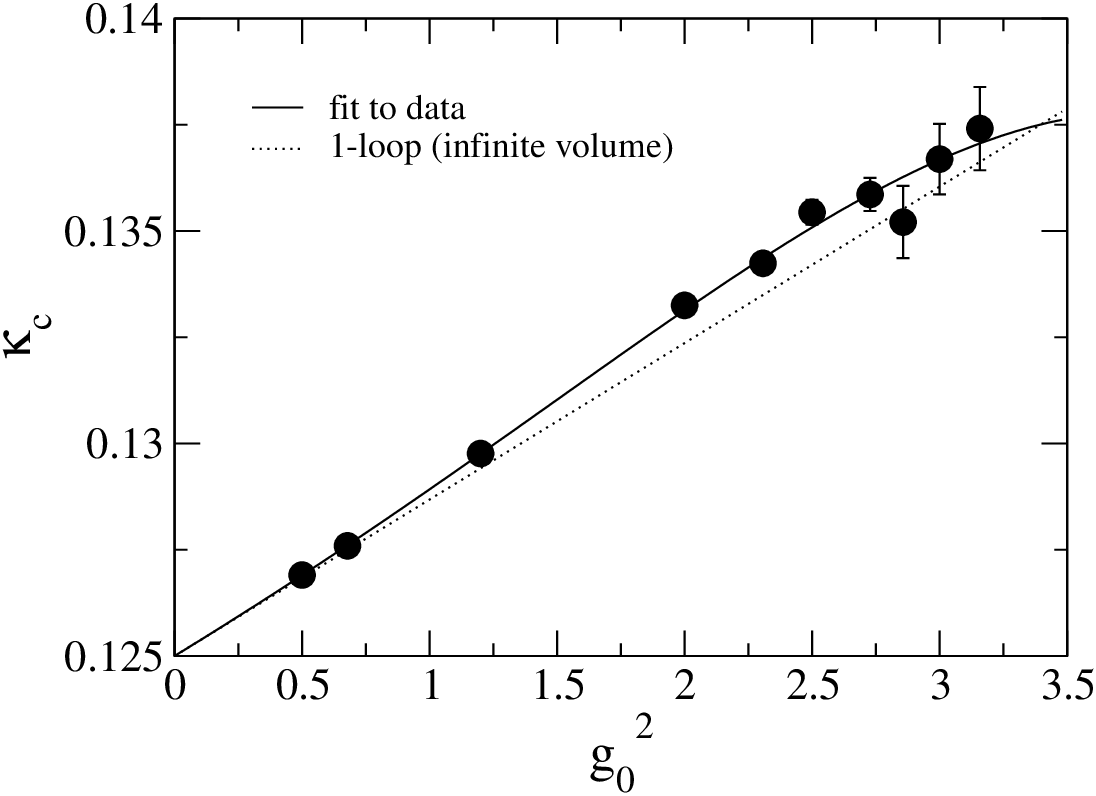}\\
\includegraphics[scale=\figscale,clip=]{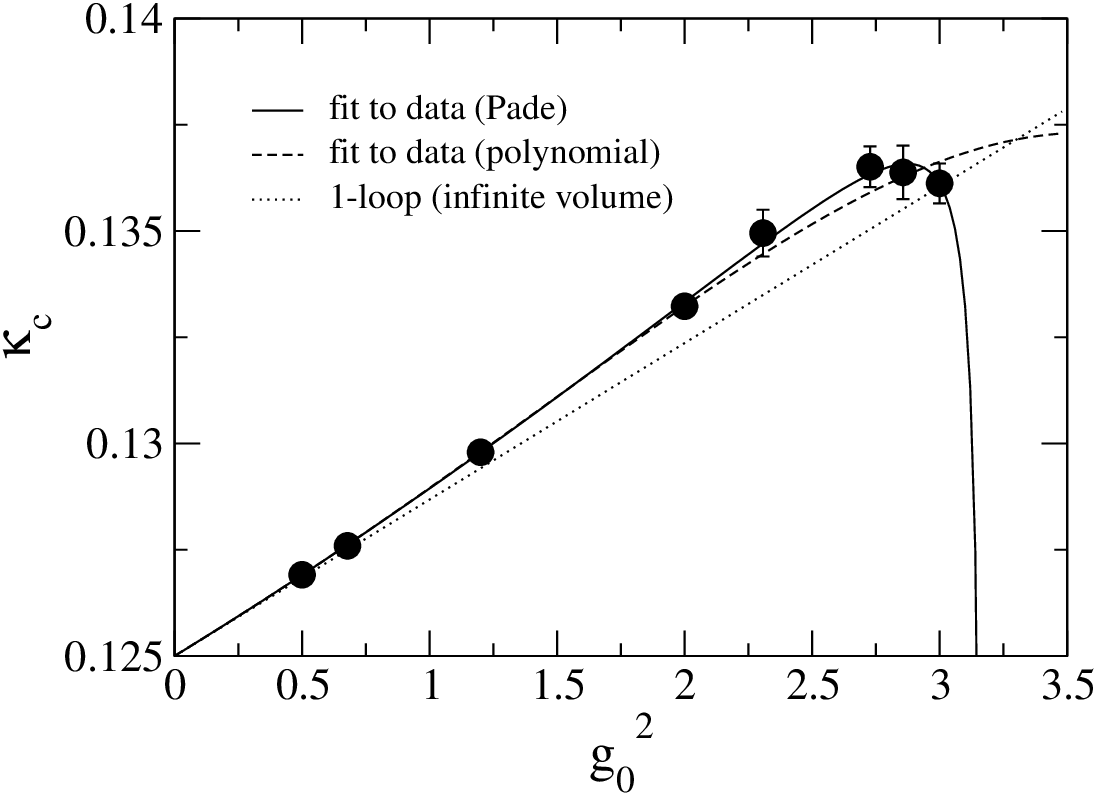}\\
\includegraphics[scale=\figscale,clip=]{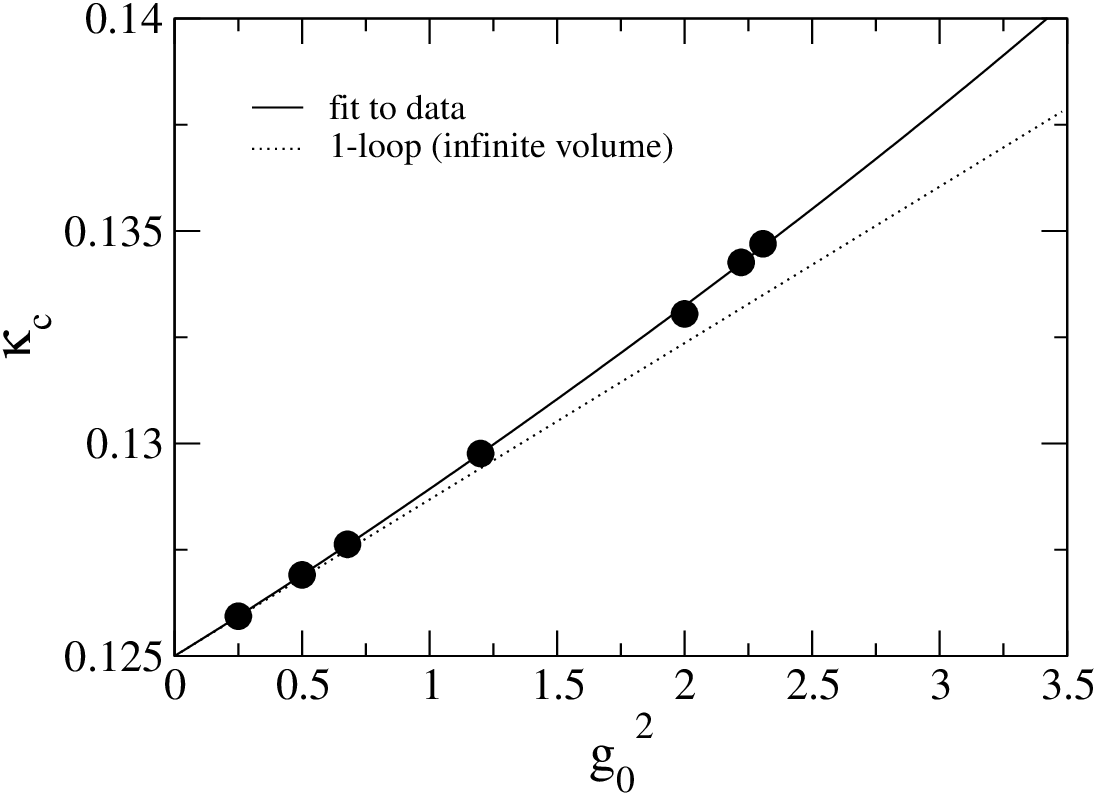}
\label{fig:Nf320KappaC}
\end{figure}
\end{widetext}

\end{document}